\documentclass[a4paper]{report}
\usepackage[utf8]{inputenc}
\usepackage[T1]{fontenc}
\usepackage{RJournal}
\usepackage{amsmath,amssymb,array}
\usepackage{booktabs}

\usepackage{multirow}

\newcommand{\bX}{ \mbox{\bf X}}

\newcommand{\bM}{ \mbox{\bf M}}
\newcommand{\bW}{ \mbox{\bf W}}

\newcommand{\iid}{\stackrel{iid}{\sim}}
\newcommand{\indep}{\stackrel{indep}{\sim}}

\DeclareMathOperator*{\argmin}{arg\,min}

\newcommand{\beq}{ \begin{equation}}
\newcommand{\eeq}{ \end{equation}}
\newcommand{\beqn}{ \begin{eqnarray}}
\newcommand{\eeqn}{ \end{eqnarray}}
\newcommand{\lcbk}{\left\{}
\newcommand{\rcbk}{\right\}}
\newcommand{\lsbk}{\left[}
\newcommand{\rsbk}{\right]}
\newcommand{\lpth}{\left(}
\newcommand{\rpth}{\right)}

\begin{document}

%% do not edit, for illustration only
\sectionhead{Contributed research article}
\volume{XX}
\volnumber{YY}
\year{20ZZ}
\month{AAAA}

%% replace RJtemplate with your article
\begin{article}
  % !TeX root = RJwrapper.tex
\title{\pkg{SPQR}: An R Package for Semi-Parametric Density and Quantile Regression}
\author{by Steven G. Xu, Reetam Majumder and Brian J. Reich}

\maketitle

\abstract{
We develop an R package \pkg{SPQR} that implements the semi-parametric quantile regression (SPQR) method in \cite{xu2021bayesian}. The method begins by fitting a flexible density regression model using monotonic splines whose weights are modeled as data-dependent functions using artificial neural networks. Subsequently, estimates of conditional density and quantile process can all be obtained. Unlike many approaches to quantile regression that assume a linear model, SPQR allows for virtually any relationship between the covariates and the response distribution including non-linear effects and different effects on different quantile levels. To increase the interpretability and transparency of SPQR, model-agnostic statistics developed by \cite{ApleyZhu2020} are used to estimate and visualize the covariate effects and their relative importance on the quantile function. In this article, we detail how this framework is implemented in \pkg{SPQR} and illustrate how this package should be
used in practice through simulated and real data examples.
}

\section{Introduction}
Quantile regression (QR) \citep{KoenkerBasset1978} is a widely used tool in applications where covariate effect on non-central part of the response distribution is of interest. Such problems routinely arrive in economics, health and environment studies where the response is often heavy-tailed and/or heteroscatistic. Although modeling a single quantile or a small subset of quantiles often serves as the primary interest in most applications, the full potential of QR lies in estimation of the full quantile process to obtain a comprehensive description of the varying covariate effects across the quantile domain.

Most existing implementations of quantile regression in R assume a linear relationship between the conditional quantile and covariates, examples include \CRANpkg{bayesQR} \citep{bayesQR}, \CRANpkg{lqr} \citep{lqr} and \CRANpkg{quantreg} \cite{quantreg}. These methods become too restrictive, leading to large estimation bias, when the underlying relationship is highly nonlinear. To overcome this challenge, semi-parametric QR models that allow some or some transformation of the covariate effects be modeled flexibly have been proposed, examples are \CRANpkg{quantreg.nonpar} \citep{npqr}, \CRANpkg{plaqr} \citep{plaqr} and \CRANpkg{qgam} \citep{qgam} that use generalized additive models; and \CRANpkg{siqr} \citep{siqr} that uses a single index model. Additive models allow flexible and interpretable modeling of main effects, but explicitly specifying the interaction terms quickly become a tedious task as the data dimension increases. Single index model alleviates curse of dimensionality by projecting the covariates to a 1-dimensional space, but when the intrinsic data dimension is high such projection can be too coarse to retain all valuable information. A more serious disadvantage shared by all aforementioned methods is that they do not ensure the monotonicity of the estimated conditional quantile function. Without such constraint, the estimated quantiles can cross each other when the sample size is small, leading to a invalid response distribution that is difficult or impossible to draw inference from.

Few existing packages are capable of estimating multiple quantiles under non-crossing constraints. The \CRANpkg{qrjoint} package \citep{qrjoint} directly models the linear coefficient function as a monotone process and allows simultaneous estimation of non-crossing quantile planes. The \code{mcqrnn()} function in the \CRANpkg{qrnn} package \citep{mcqrnn} can model non-linear non-crossing quantile curves by treating the quantile level as an observed covariate and imposing positivity constraints on its correspond weights in the neural network. However, the model suffers from high variance when the sample size is small \citep{xu2021bayesian} and uncertainty quantification is not straightforward.

The purpose of this article is to introduce a new R package, named \CRANpkg{SPQR} \citep{SPQR}, for flexible QR modeling. This package implements the semi-parametric quantile regression (SPQR) model proposed by \citet{xu2021bayesian} which allows simultaneous estimation of non-linear non-crossing quantile curves. The method begins by specifying a semi-parametric model for the conditional distribution function using shape-constrained splines, the coefficients are then modeled non-parametrically as functions of covariates using artificial neural networks. As a result, valid estimates of the conditional response distribution and its quantiles can be simultaneously obtained. The \CRANpkg{SPQR} provides three approaches for fitting SPQR: maximum likelihood estimation (MLE) and maximum \emph{a posteriori} probability (MAP) which provide point estimates, and Markov chain Monte Carlo (MCMC) which provides uncertainty quantification through posterior samples. To our best knowledge and at the time of writing, \pkg{SPQR} is the only package that implements a fully Bayesian framework for estimating semi-parametric non-crossing QR. The main computations of \pkg{SPQR} rely on \CRANpkg{torch} \citep{torch} and \CRANpkg{Rcpp} \citep{Rcpp}. Specifically, MLE and MAP are optimized using the Adam routine \citep{kingma2014} in \pkg{torch} and take advantage of its GPU capability, and MCMC is implemented using \pkg{Rcpp} and \CRANpkg{RcppArmadillo} \citep{RcppArmadillo} for efficient gradient computation. To increase the interpretability and transparency of SPQR, \pkg{SPQR} also provides function to compute quantile accumulative local effects \citep[ALE;][]{ApleyZhu2020} to characterize and visualize quantile-dependent main and interaction effects, as well as function to compute quantile-dependent variable importance measures.

The rest of the article is organized as follows. We first give a brief review of the methodology background of SPQR and quantile ALEs. We then introduce the implemented computation approaches for fitting SPQR. The subsequent section introduces the \pkg{SPQR} package and its main features in detail. Finally, we demonstrate the usage and effectiveness of \pkg{SPQR} in density and quantile regression problems using simulated and real data. In particular, we apply our package to the Australia electric demand data, available in the \pkg{qgam} package, and analyze the quantile effects of temperature and time on residential electric energy consumption. The last section concludes the paper.

\section{Methodology background}
\subsection{Density regression model}
To estimate quantile effects, we first build a model for the probability density function (PDF) of the response $Y$ conditioned on covariates $\bX=(X_1,...,X_p)$.  We assume that the response and all $p$ covariates are scaled to $[0,1]$ to simplify the specifications of prior distributions and basis functions.  The PDF is assumed to be a function of $K$ second-order M-spline basis functions, $M_1(y),...,M_K(y)$, with equally-spaced knots spanning $[0,1]$.  Each M-spline basis function is itself a valid PDF on $[0,1]$ \citep{ramsay1988}, and therefore any convex combination of these functions is also a valid PDF.  The model is 
\begin{equation}
 f(y|\bX) = \sum_{k=1}^K\theta_k(\bX)M_k(y),
\end{equation}
where the probabilities $\theta_k(\bX)$ satisfy $\theta_k(\bX)\ge 0$ and $\sum_{k=1}^K\theta_k(\bX)=1$ for all possible $\bX$.  

To provide flexibility and ensure non-negative weights $\theta_k(\bX)$ that sum to one, we use a fully connected neural network (NN) with softmax output activation.  An NN with $L$ layers ($L-1$ hidden layers) is parameterized by a set of weight matrices $\mathcal{W}=\{\bW^{(1)}, ...,\bW^{(L)}\}$, with $\bW^{(l)}\in\mathbb{R}^{(V_{l}+1)\times V_{l-1}}$ where $V_l$ is the number of units (excluding the intercept/bias node) in layer $l$. We define the input layer as layer $0$ for simplicity. %Let $\hat{\theta}_k(\bX,\mathcal{W})$ denote the estimator of $\theta_k(\bX)$ based on unknown parameters $\mathcal{W}$, the NN maps the input $\bX$ to the probabilities $\hat{\theta}_k(\bX,\mathcal{W})$ through three steps.
The first layer is
\begin{equation}
    z^{(1)}_i(\bX,\mathcal{W}) = W^{(1)}_{i0} + \sum_{j=1}^pW^{(1)}_{ij}X_j .
\end{equation}
We make the functional dependence on $\bX$ and $\mathcal{W}$ explicit in our notation as it will help clarify what
follows. Subsequent hidden layers are defined by the recursion
\begin{equation}\label{e:rec}
    \begin{split}
    u_i^{(l)}(\bX,\mathcal{W}) &= \phi\lcbk z^{(l)}_i(\bX,\mathcal{W})\rcbk\mbox{\ \ \ for\ \ } l\in\{1,...,L-1\} \\
    z^{(l+1)}_i(\bX,\mathcal{W}) &= W_{i0}^{(l+1)} + \sum_{j=1}^{V_l}W^{(l+1)}_{ij}u_j^{(l)}(\bX,\mathcal{W})
    \end{split}
\end{equation}
where $\phi$ is the activation function taken to be either the hyperbolic tangent function $\phi(u) = (e^{2u}-1)/(e^{2u}+1)$ or the rectified-linear function $\phi(u) = \max(0,u)$. Finally, the softmax activation is used in the output layer to ensure probabilities sum to one
\begin{equation}\label{e:FNN}
     \theta_k(\bX,\mathcal{W}) = \frac{\exp\lcbk z^{(L)}_k(\bX,\mathcal{W})\rcbk}{\sum_{i=1}^{K}\exp\lcbk z^{(L)}_i(\bX,\mathcal{W})\rcbk}.
\end{equation}
The SPQR model for the conditional PDF is then
\begin{equation}\label{e:PDF}
    f(y|\mathcal{W},\bX)=\sum_{k=1}^KM_k(y)\frac{\exp\lcbk z^{(L)}_k(\bX,\mathcal{W})\rcbk}{\sum_{i=1}^{K}\exp\lcbk z^{(L)}_i(\bX,\mathcal{W})\rcbk}.
\end{equation}
Endowed with the approximation theories of spline and NN, the model in \eqref{e:PDF} can approximate any smooth PDF as $K$ and the $V_l$ increase. Therefore, this model can capture complex relationships, such as covariate-dependent variance, skewness, or likelihood of extreme events.

Computing the PDF is simple and fast given the parameters $\mathcal{W}$, and since the integral of M-spline functions are I-spline functions \citep{ramsay1988}, \eqref{e:PDF} immediately gives rise to an expression for the cumulative distribution function (CDF) 
\begin{equation}\label{e:CDF}
 F(y|\mathcal{W},\bX) = \sum_{k=1}^KI_k(y)\frac{\exp\lcbk z^{(L)}_k(\bX,\mathcal{W})\rcbk}{\sum_{i=1}^{K}\exp\lcbk z^{(L)}_i(\bX,\mathcal{W})\rcbk},
\end{equation}
where $I_k(y)$ are I-spline basis functions. The conditional quantile function for quantile level $\tau\in(0,1)$ is defined as the function $Q(\tau|\mathcal{W},\bX)$  so that 
$$F\{Q(\tau|\mathcal{W},\bX)|\mathcal{W},\bX\}=\tau.$$ The conditional quantile function for this model is not available in closed-form, but can be approximated by numerically inverting $F(y|\mathcal{W},\bX)$. Given that \eqref{e:CDF} models a valid CDF, the conditional quantile function estimated through this approach satisfies the non-crossing constraint $$\frac{\partial Q(\tau|\mathcal{W},\bX)}{\partial\tau}>0,\ \forall\ \bX$$ and does not require any second-stage monotonization treatment. 

\subsection{Summarizing covariate effects on quantiles}
SPQR has the advantage of being a flexible semi-parametric model that can capture complex non-linear covariate effects on various aspects of the response distribution.  A disadvantage is that it is difficult to interpret individual parameters because the weights $\mathcal{W}$ are not individually identified and do not correspond to meaningful quantities. In most applications where quantile regression is used, understanding the covariate effect on different quantiles is of paramount interest. Therefore, we seek to quantify covariate effects on specific aspects of the response distribution as measured by the quantile function $Q(\tau|\bX)$.

We quantify covariate quantile effects using the accumulative local effects (ALEs) of \cite{ApleyZhu2020}. The sensitivity of $Q(\tau|\mathcal{W},\bX)$ to covariate $j$ is naturally quantified by the partial derivative $$q_j(\tau|\mathcal{W},\bX) = \frac{\partial Q(\tau|\mathcal{W},\bX)}{\partial X_j}.$$ The ALE begins by averaging $q_j(\tau|\mathcal{W},\bX)$ over $\bX$ conditioned on $X_j = u$,  
\begin{equation*}
    {\bar q}_j(\tau|\mathcal{W},u) = \mathbb{E}_{\bX}\{ q_j(\tau|\mathcal{W},\bX)|X_j=u\}.
\end{equation*}
The ALE main effect function of $X_j$ is then defined as 
\begin{equation*}
    ALE_j(\tau|\mathcal{W},x) =\int_{0}^{x}{\bar q}_j(\tau|\mathcal{W},u)du.
\end{equation*}
%When the ALE function is estimated for a model that predicts the conditional mean, as in its original proposal \citep{ApleyZhu2020}, it is often centered to have mean 0 with respect to the marginal distribution of $X_j$ so that it has the interpretation of additional effect brought by $X_j$ on the estimator $\hat{\mathbb{E}}(Y|X_{-j})$ where $X_{-j}$ is the covariate vector except $X_j$. This property does not carry over to the quantile regression setting, so we just define the quantile ALE effect of $X_j$ as $\mbox{ALE}_j(\tau|\mathcal{W},x_j)={\bar Q}_j(\tau|\mathcal{W},x)$. 
Analogous formulas define the second-order ALE interaction effect for $X_j$ and $X_l$, $\mbox{ALE}_{jl}(\tau|\mathcal{W},x_j,x_l)$, by taking the partial derivative with respect to both $X_j$ and $X_l$. These functions can be plotted by $\tau$ to summarize how the predicted quantile changes with respect to change in the covariate values. In addition to the ALEs, we follow \cite{greenwell2018} and distill the ALE function to one-number summaries to compare variable importance by quantile level. The variable importance (VI) for continuous covariates are characterized by the standard deviation of the ALE with respect to the marginal distribution of $\bX$, i.e., $\mbox{VI}_j(\tau|\mathcal{W}) = \mbox{SD}\{ ALE_j(\tau|\mathcal{W},X_j)\}$ and $ \mbox{VI}_{jl}(\tau|\mathcal{W}) = \mbox{SD}\{ ALE_{jl}(\tau|\mathcal{W},X_j,X_l)\}$. For discrete covariates with few unique levels, the standard deviation is replaced by the range. 

The ALE and VI summaries depend on the model parameters, $\mathcal{W}$. They can either be evaluated using a point-estimate $\widehat{\mathcal{W}}$ to give a point-estimate of the summaries, $ALE_j(\tau|\widehat{\mathcal{W}},x)$ and $\mbox{VI}_j(\tau|\widehat{\mathcal{W}})$, or, for a Bayesian analysis, the posterior samples can be used to quantify uncertainty of the summaries such as the posterior probability that variable $j$ is more important than variable $l$ for predicting conditional quantile at $\tau$.

\section{Computational approaches}\label{s:computing}

\pkg{SPQR} includes four computational algorithms to estimate the model parameters and importance measures: maximum likelihood estimation (MLE), maximum \emph{a posteriori} probability (MAP), Hamiltonian Monte Carlo (HMC) and the no-U-turn sampler (NUTS). These algorithms are described in detail below. 

\subsection{Maximum likelihood estimation (MLE)}\label{s:MLE}

Directly modeling the conditional PDF means that the negative likelihood function given $n$ training observations $(\bX_i,y_i)$ for $i\in\{1,...,n\}$ has a closed form,
\begin{equation}\label{e:NLL}
    \ell(\mathcal{W}) = -\sum_{i=1}^n\log\lcbk\sum_{k=1}^KM_k(y_i)\theta_{k}(\bX_i,\mathcal{W})\rcbk
\end{equation}
and can be used as a loss function to be minimized, leading to the MLE estimator
\begin{equation}\label{e:mle}
    \widehat{\mathcal{W}}_{\text{MLE}} = \argmin_\mathcal{W} \ell(\mathcal{W}).
\end{equation}
Equation \eqref{e:mle} can be solved using standard back-propagation algorithms. The MLE estimator,
however, does not put any constraint on the magnitude of $\mathcal{W}$ and therefore can result in an unstable model that overfits the data. One solution is to augment the loss function in \eqref{e:NLL} with a regularization term that penalizes large $\mathcal{W}$, for example the \textit{weight decay} regularization. However, carefully choosing the penalty coefficients is crucial to the predictive performance but increases computation complexity. Furthermore, uncertainty analysis of the estimators obtained by \eqref{e:mle} is not straightforward and has to rely on bootstrap approaches, bringing additional computational challenges to the already complex problem.

\subsection{Bayesian estimation}
To address the lack of uncertainty quantification posed by MLE, we adopt a Bayesian framework and give prior distribution for the weight parameters. The uncertainty of the SPQR estimators is then characterized by their posterior distributions. Specifying prior distributions also provides regularization to the model and stabilizes weight estimation. We assume hierarchical normal priors for the weights 
\begin{equation}\label{e:priors}
    W_{ij}^{(l)}|\sigma^{(l)},\lambda_{j}^{(l)} \sim \mathcal{N}\left(0,\sigma^{(l)2}\lambda_{j}^{(l)2}\right), \ p(\lambda_j^{(l)})\sim p(\lambda_j^{(l)};\gamma_{\lambda})\mbox{\ \ \ for\ \ } j\ge 0
\end{equation}
where $\sigma^{(l)}$ is the layer-wise global scale shared by all weights in layer $l$, which can either
be set to a constant value or estimated using non-informative priors, and $\lambda_{j}^{(l)}$ is a unit-wise local scale with hyper-prior $p(\lambda_j^{(l)};\gamma_\lambda)$. Hierarchical normal distribution is the most commonly used prior for NN weights as they impose a \textit{weight decay} penalty with estimable penalty coefficients, and many priors proposed in the Bayesian NN literature are variants of \eqref{e:priors}. In \pkg{SPQR}, we consider three of such models as summarized in Table \ref{t:priors}: the Gaussian Process (GP) prior, the Automatic Relevance Determinantion (ARD) prior, and the Gaussian Scale Mixture (GSM) prior. 

The GP prior is proposed by \citet{neal1996} who considers the weights and bias in each layer as separate parameter blocks. The weights in each layer depend on a common variance, $\lambda^{(l)}$, and the bias is given a separate variance. In addition, the variances on weights are scaled by the width of the layer, $V_{l-1}$, for all but the input layer. The NN under such setting is shown to converge to a certain multi-output Gaussian process under the condition that $V_l\rightarrow\infty$ for $l\in[1,L-1]$ in both the case of $L=2$ \citep{neal1996} and $L\rightarrow\infty$ \citep{matthews2018}. The GP prior assumes that the weights marginally follow Gaussian distributions, since they share a common variance hyperparameter. Recent studies on Bayesian NNs, however, have found that distribution of weights in a deep NN (DNN) can be heavy-tailed \citep{vladimirova2019,fortuin2021}. Therefore, it might be helpful to incorporate such knowledge and allow a wider model coverage for the prior distributions for weights. The ARD prior was originally developed by \citet{mackay1992} who assigns weights in the input-to-hidden layer unit-wise local scales so that the magnitude of weights associated with each input will determine its relevance. Under this setting, the marginal distribution of the input weights, $W_{ij}^{(1)}|\sigma^{(1)}$, depends on the hyper-prior for local scales, $\lambda^{(1)}_j$, through the integration
\begin{equation}\label{e:gsm}
    p(w_{ij}^{(1)}|\sigma^{(1)}) = \int\mathcal{N}(0,\sigma^{(1)2}\lambda_{j}^{(1)2})p(\lambda_{j}^{(1)};\gamma_\lambda)d\lambda_{j}^{(1)}.
\end{equation}
For all other layers, the ARD prior has the same structure as that of the GP prior. The GSM prior is a direct generalization of the ARD prior by allowing the layer-wise global scale $\sigma^{(l)}$ to be estimable and all layers to have the flexibility of \eqref{e:gsm}. Therefore, it not only determines the relevance of each input feature but also that of each latent feature in deeper layers.

\begin{table}[htbp]
\begin{center}
\caption{{\bf Prior distributions}: \pkg{SPQR} allows for several models for the prior distribution for the layer-wise global scale,  $\sigma^{(l)}$, and unit-wise local scale, $\lambda_j^{(l)}$. This table gives the Gaussian Process (GP), Automatic Relevance Determination (ARD) and Gaussian Scale Mixture (GSM) priors in terms of unit index in layer $l$, $j\in[0,V_l]$, and hyperparameters $\gamma_\sigma$ and $\gamma_\lambda$.}\label{t:priors}
\begin{tabular}{cccc}
\toprule
Name of the prior & $\sigma^{(l)}$ & $\lambda^{(l)}_0$ & $\lambda^{(l)}_j,\ \forall j\ge1$ \\
\midrule
\multirow{3}{*}{GP} & \multirow{3}{*}{1} & \multirow{3}{*}{$p(\lambda_0^{(l)};\gamma_{\lambda})$} & $\lambda^{(l)}\sim p(\lambda^{(l)};\gamma_{\lambda})$\\
& & & $\lambda_j^{(1)}=\lambda^{(1)};\ \lambda_j^{(l)}=\lambda^{(l)}/V_{l-1},\ \forall l\ge 2$\\
\midrule
\multirow{2}{*}{ARD} & \multirow{2}{*}{1} & \multirow{2}{*}{$p(\lambda_0^{(l)};\gamma_{\lambda})$} & $\lambda_j^{(1)}\iid p(\lambda_j^{(1)};\gamma_{\lambda})$\\
& & &  $\lambda_j^{(l)}\sim \mbox{GP},\ \forall l\ge 2$\\
\midrule
\multirow{2}{*}{GSM} & \multirow{2}{*}{$p(\sigma^{(l)};\gamma_{\sigma})$} & \multirow{2}{*}{$p(\lambda_0^{(l)};\gamma_{\lambda})$} & \multirow{2}{*}{$\lambda_j^{(l)}\iid p(\lambda_j^{(l)};\gamma_{\lambda})$}\\
& & & \\
\bottomrule
\end{tabular}
\end{center}
\end{table}

To complete the prior specification, we assign the local scale, $\lambda_j^{(l)}$, non-informative hyper-priors. In the case of GSM, we also specify a non-informative hyper-prior for the global scale, $\sigma^{(l)}$. A common choice is the inverse-Gamma distribution,
\begin{equation}
\begin{split}\label{e:igamma}
    p(\cdot;\gamma_{\lambda})&=\mathcal{IG}amma(a_{\lambda},b_{\lambda})\\
    p(\cdot;\gamma_{\sigma})&=\mathcal{IG}amma(a_{\sigma},b_{\sigma}),
\end{split}
\end{equation}
which is also what will be assumed in this article. Under this setting, the marginal distribution of weights $p(w_{ij}^{(l)}|\sigma^{(l)})$ follows a Student-t whose degrees-of-freedom is determined by the hyperparameters, $a_{\lambda}$ and $b_{\lambda}$. Notice that in the cases of GP prior, by the scaling property of Gamma distributions, the prior for $\lambda_j^{(l)},\ l\ge 2$ is also inverse-Gamma
\begin{equation}
    \lambda_j^{(l)}\indep \mathcal{IG}amma(a_{\sigma}, \frac{b_{\sigma}}{V_{l-1}})\mbox{\ \ \ for\ \ } j\ge 2
\end{equation}

\subsubsection{Maximum {\it a posteriori} estimation (MAP)}\label{s:MAP}

Before we introduce the fully Bayesian approach to estimate NN parameters using HMC, we note that the prior specification in \eqref{e:gsm} enables the adoption of a MAP method to estimate the weights, i.e.,
\begin{equation}\label{e:map}
    \begin{split}
        \widehat{\mathcal{W}}_{\text{MAP}},\ \hat{\boldsymbol{\sigma}},\ \hat{\boldsymbol{\Lambda}}&=\argmin_{\mathcal{W}, \boldsymbol{\sigma}, \boldsymbol{\Lambda}}\lcbk\ell(\mathcal{W})-\log p(\mathcal{W},\boldsymbol{\sigma},\boldsymbol{\Lambda})\rcbk\\
        p(\mathcal{W},\boldsymbol{\sigma},\boldsymbol{\Lambda})&=\prod_{k=1}^K\prod_{l=1}^L\prod_{j=0}^{V_l} \mathcal{N}(W_{kj}^{(l)}|0,\sigma^{(l)}\lambda_{j}^{(l)})p(\sigma^{(l)}|\gamma_{\sigma})p(\lambda_j^{(l)}|\gamma_{\lambda}).
    \end{split}
\end{equation}
Solving the optimization problem in \eqref{e:map} will not lead to sensible results as the gradients of $\sigma^{(l)}$ and $\lambda_{j}^{(l)}$ do not depend on the data. We adopt a reparameterization of \eqref{e:map} to let the likelihood function become direct function of the scale hyperparameters
\begin{equation}\label{e:ncp}
    \begin{split}
        \widehat{\mathcal{Z}},\ \hat{\boldsymbol{\sigma}},\ \hat{\boldsymbol{\Lambda}}&=\argmin_{\mathcal{Z}, \boldsymbol{\sigma}, \boldsymbol{\Lambda}}\lcbk\ell(\mathcal{W})-\log p(\mathcal{Z},\boldsymbol{\sigma},\boldsymbol{\Lambda})\rcbk\\
        W_{kj}^{(l)}&=\sigma^{(l)}\lambda_j^{(l)}Z^{(l)}_{kj}\\
        p(\mathcal{Z},\boldsymbol{\sigma},\boldsymbol{\Lambda})&=\prod_{k=1}^K\prod_{l=1}^L\prod_{j=0}^{V_l} \mathcal{N}(Z_{kj}^{(l)}|0,1)p(\sigma^{(l)}|\gamma_{\sigma})p(\lambda_j^{(l)}|\gamma_{\lambda}).
    \end{split}
\end{equation}
Equation \eqref{e:ncp} can be solved by substituting the prior distributions for variance hyperparameters with any of the three models described in Table~\ref{t:priors}. The MAP estimates of $\mathcal{W}$ can then be calculated as $\widehat{W}_{kj,\text{MAP}}^{(l)}=\hat{\sigma}^{(l)}\lambda_j^{(l)}\hat{Z}^{(l)}_{kj}$. As with the MLE method, MAP estimation uses standard back-propagation algorithms, provides only a point estimate of ${\cal W}$ and does not allow for direct quantification of model uncertainty.

\subsubsection{Hamiltonian Monte Carlo (HMC)}\label{s:HMC}

MCMC produces samples from the posterior distribution of the parameters that can be used to approximate their entire posterior distribution. We follow the strategy of \citet{neal1996} and use a block-updating scheme that utilizes two MCMC algorithms. The conditional distribution of weights given the variance hyperparameters and the data are approximated using the Hamiltonian Monte Carlo (HMC) sampler, whereas the conditional distribution of variance hyperparameters given the weights are approximated using the Gibbs sampler \citep{geman1984}. HMC permits efficient sampling from a high-dimensional target distribution by using its gradient with respect to each direction. It uses an approximate Hamiltonian dynamics simulation based on numerical integration which allows the sampler to explore more carefully in regions with high density and escape quickly from regions with low density \citep{betancourt2017}. The candidate value of $\mathcal{W}$ found by this simulation is then accepted with a Metropolis step to correct for any numerical error resulted from the numerical integration. 

We consider two implementations of the HMC sampler: the vanilla HMC sampler \citep{neal2011} and the more advanced no-U-turn sampling \citep[NUTS;][]{hoffman2014}. The vanilla HMC sampler requires setting the number of approximate integration time $t$, which is the number of leap-frog steps $L_{\epsilon}$ multiplied by the step size $\epsilon$. The step size is automatically optimized during warmup sample iterations using dual-averaging to match an acceptance-rate target, and the number of steps will be calculated as $L_{\epsilon}=\lfloor t/\epsilon\rfloor$. An Euclidean metric $\bM$, where $\bM^{-1}$ estimates the posterior covariance of $\mathcal{W}$, is also estimated during warmup to help project the parameters to a space where sampling can be done more efficiently. The NUTS, compared to HMC, has the further advantage of adaptively setting the number of leap-frog steps $L_{\epsilon}$ on the fly during both warmup and sampling. This greatly reduces the required effort on the users' side to select a reasonable value for the number of approximate integration time $t$.

Let $\mathcal{W}_s$ be the posterior samples of $\mathcal{W}$ after discarding warmup iterations, for $s\in\{1,...,S\}$. Since the NN is over-parameterized and individual weights $W^{(l)}_{ij}\in\mathcal{W}$ are usually unidentified, the posterior distribution of the weights themselves might not be very meaningful. However, the samples can produce estimates of meaningful quantities such as the conditional quantile function (QF) $Q(\tau|\bX,\mathcal{W})$. For example, the conditional QF estimator is the posterior mean
\begin{equation*}
    \hat{Q}_{\text{MCMC}}(\tau|\bX,\mathcal{W}) = \frac{1}{S}\sum_{s=1}^SQ(\tau|\bX,\mathcal{W}_s)
\end{equation*}
and point-wise credible bands are obtained as the sample quantiles of $Q(\tau|\bX,\mathcal{W}_s)$.
\begin{table}[htbp]
\begin{center}
\caption{The overview of functions in package \pkg{SPQR}.}\label{t:SPQR}
\setlength\extrarowheight{2pt} % for a bit of visual "breathing space"
\begin{tabular}{p{0.3\textwidth}p{0.64\textwidth}}
\toprule
Function & Description \\
\midrule
\code{SPQR()}& Main function of the package. Fits SPQR using the MLE, MAP, or MCMC method. Returns an object of S3 class \code{"SPQR"}, a list which includes the fitted model (\code{model}), the model configuration (\code{config}), the control parameters (\code{control}), the running time (\code{time}), the covariate matrix (\code{X}), the response vector (\code{Y}), as well as method-dependent training information. \\
\code{cv.SPQR()}& Fits SPQR using MLE or MAP method, and computes K-fold cross-validation (CV) error.\\
\code{createFolds.SPQR()}& Generate pre-computed CV folds.\\
\code{summary()}& Extracts and computes a list of summary information of a \code{"SPQR"} class object. Returns an object of S3 class \code{"summary.SPQR"}.\\
\code{print.summary()}& Prints the contents of a \code{"summary.SPQR"} class object in a user-friendly way.\\
\code{print()}& Computes and prints the summary information of a \code{"SPQR"} class object. Equivalent to \code{print.summary(summary())}.\\
\code{coef()}& Computes and returns the estimated spline coefficients $\theta_k(\bX,\widehat{\mathcal{W}})$ of a \code{"SPQR"} class object.\\
\code{predict()}& Computes and returns the estimated PDF/CDF/QF of a \code{"SPQR"} class object.\\
\code{QALE()}& Computes and returns the quantile accumulative local effects (ALE) of a \code{"SPQR"} class object.\\
\code{plotEstimator()}& Computes and plots the estimated PDF/CDF/QF curves of a \code{"SPQR"} class object.\\
\code{plotGOF()}& Performs a visual goodness-of-fit test for the estimated conditional PDF using probability inverse transformation method.\\
\code{plotMCMCtrace()} & Show trace plot of the log-likelihood or a specified estimate of a \code{"SPQR"} class object fitted with \code{method="MCMC"}.\\
\code{plotQALE()}& Computes and plots the quantile ALE effects of a \code{"SPQR"} class object.\\
\code{plotQVI()}& Computes and plots the ALE-induced quantile variable importance measures of a \code{"SPQR"} class object.\\
\code{autoplot()}& A wrapper function that creates a user-specified plot for a \code{"SPQR"} class object by calling one of the plot functions above.\\
\bottomrule
\end{tabular}
\end{center}
\end{table}

In summary, the MLE and MAP approaches estimate the parameters by directly minimizing a loss function and are computationally faster, but they only provide point estimates of $\mathcal{W}$ as well as the ALE and VI metrics. HMC is a Markov chain Monte Carlo (MCMC) algorithm that approximates the posterior distribution of the parameters and is computationally slower, but produces posterior samples that can be used to quantify uncertainty. However, HMC requires careful tuning to ensure sampling efficiency. NUTS is an adaptive variant of HMC that requires little tuning at the further expense of computational efficiency. Therefore, we recommend the MLE and MAP estimator for huge datasets where uncertainty quantification is less of a concern, NUTS for small datasets where uncertainty quantification is critical, and HMC as a balance between these options.

\section{The \pkg{SPQR} package}

The R package \pkg{SPQR} has two model fitting functions, \code{SPQR()} and \code{cv.SPQR()}, as well as various helper functions for handling tasks such as model validation, model prediction, and results visualization. Table~\ref{t:SPQR} lists all the main functions provided by \pkg{SPQR}.

\subsection{The main fitting function}

The \code{SPQR} function specifies a semi-parametric conditional density regression model of the type \eqref{e:PDF} and estimates the model parameters using one of the four computational approaches described in the previous section. It has the following arguments
\begin{example}
  SPQR(X, Y, n.knots=10, n.hidden=c(10), activation=c("tanh","relu"),
    method=c("MLE","MAP","MCMC"), prior=c("GP","ARD","GSM"), hyperpar=list(), 
    control=list(), normalize=FALSE, verbose=TRUE, seed=NULL, ...)
\end{example}
This function takes two required arguments, a $n\times p$ covariate matrix (without intercept column) \code{X} and a response vector \code{Y}. The covariate matrix is expected to contain only numeric features, and that all categorical features are converted to numeric values in advance. The covariates are also recommended, although not required, to be normalized/standardized to have the same scale to stabilize gradient based optimization that will be used to estimate the parameters. The response vector, on the other hand, is required to take values between 0 and 1. We provide a \code{normalize} argument such that, when setting \code{normalize=TRUE}, all variables will be scaled to the unit interval using min-max transformation. Their original scales will be recorded to back-transform the estimated density and quantile function. By default, however, we set \code{normalize=FALSE} as we want the users to have full control on how variables are scaled, such as using prior information on the domains of the variables, etc. 

The arguments \code{n.knots} and \code{n.hidden} are the number of basis functions, $K$, and the number of hidden neurons, $V_l$, that define the SPQR model in \eqref{e:PDF}. \pkg{SPQR} uses the function \code{mSpline()} in the \CRANpkg{splines2} package \citep{splines2} to construct the basis functions. We require setting \code{n.knots} to at least 5 as the model may severely underfit otherwise. The \code{n.hidden} argument accepts a vector of integers such that \code{n.hidden[l]} is the the number of hidden neurons in layer $l$ for $l\in\{1,...,L-1\}$. It should be noted that although \code{n.knots} and \code{n.hidden} are given default values for convenience, they should be tuned per application for optimal performance of the SPQR estimators. The argument \code{activation} corresponds to the hidden layer activation function $\phi$ in \eqref{e:rec}, and is set to \code{activation="tanh"} for hyperbolic tangent by default. 

The argument \code{method} determines the computational approach to be used to estimate the model parameters $\mathcal{W}$. The MLE and MAP estimators are obtained by solving \eqref{e:mle} and \eqref{e:map} respectively using gradient-based stochastic optimization. Specifically, we use the Adam optimizer in the \pkg{torch} package. The \pkg{torch} package is an R implementation of the open source machine learning platform -- PyTorch \citep{pytorch}. It supports hardware acceleration for systems with a CUDA-compatible NVIDIA graphical processing unit (GPU), which \pkg{SPQR} takes advantage of. The HMC and NUTS algorithms are implemented using C++ and mirror those in Stan \citep{stan}. We did not use Stan directly for two reasons. First, Stan uses automatic differentiation whereas the analytical gradients of NNs are fairly easy to derive and evaluate. \pkg{SPQR} depends on \pkg{Rcpp} and \pkg{RcppArmadillo} to efficiently compute the log-posterior and its gradients. Secondly, Stan does not allow block-updating using both HMC and Gibbs sampler that exploits the conjugacy of hyper-priors. By default, we set \code{method="MAP"} since it is computationally faster than the MCMC method and less prone to overfit than the MLE method.

The argument \code{prior} is used only for the Bayesian methods and corresponds to one of the three variance hyperpriors described in Table~\ref{t:priors}. We set \code{prior="GP"} as default for simplicity. The argument \code{hyperpar} is a list of named hyper-prior hyperparameters to use instead of the default
values, including \code{a\_lambda}, \code{b\_lambda}, \code{a\_sigma} and \code{b\_sigma}. They correspond to $a_\lambda$, $b_\lambda$, $a_\sigma$ and $b_\sigma$ in \eqref{e:igamma}. The default value is 0.001 for all four hyperparameters. 

The argument \code{control} is a list of named and method-dependent parameters that allows finer control of the behavior of the computational approaches. The available parameters for MLE and MAP estimators are shown in Table~\ref{t:adam}. The NNs for MLE and MAP estimators are structured using a templated module that consists of only fully-connected layers (\code{"nn\_linear"}), batch normalization (\code{"nn\_batch\_norm1d"}), and dropout (\code{"nn\_dropout"}). Dropout and batch normalization are not used by default, but may be useful when training deep and wide NNs. We recommend setting \code{use.GPU=TRUE} to further accelerate computation on CUDA-configured systems. Early stopping is implemented to avoid overfitting. We use \code{valid.pct}$\times$100\% of the data as the validation set. During each epoch, a snapshot of the trained model is saved in \code{save.path/save.name} if it leads to a decrease in the validation loss. When the validation loss stops decreasing for \code{early.stopping.epochs}, the training stops and the best model is loaded and returned. 

\begin{table}[htbp]
\begin{center}
\caption{\textbf{Control parameters for MLE and MAP}.}\label{t:adam}
\setlength\extrarowheight{2pt} % for a bit of visual "breathing space"
\begin{tabular}{p{0.3\textwidth}p{0.64\textwidth}}
\toprule
Parameter & Description \\
\midrule
\code{use.GPU}& A Boolean flag for GPU utilization. Default is FALSE.\\
\code{lr}& Learning rate of Adam optimizer. Default is 0.01.\\
\code{dropout}& Dropout probabilities. A length-two vector of which the first entry specifies the dropout probability in the input-to-hidden layer and the second entry specifies the probabilities in all hidden-to-hidden layers. Default is \code{c(0,0)} which corresponds to no dropout.\\
\code{batchnorm}& A Boolean flag for batch normalization. Default is \code{FALSE}.\\
\code{epochs}& The (maximum) number of passes of the entire training dataset by Adam. Default is 200.\\
\code{batch.size}& Size of mini batches to calculate gradient. Default is 128.\\
\code{valid.pct}& Percentage of data used as validation set. Default is 0.2.\\
\code{early.stopping.epochs}& The number of epochs before stopping if the validation loss does not decrease. Default is 10.\\
\code{print.every.epochs}& The number of epochs before next training progress in printed. Default is 10.\\
\code{save.path}& Path to save the trained torch model with the lowest validation loss. Default is \code{file.path(getwd(),"SPQR\_model")}.\\
\code{save.name}& File name to save the trained \code{torch} model with the lowest validation loss. Default is \code{"SPQR.model.pt"}.\\
\bottomrule
\end{tabular}
\end{center}
\end{table}

The available parameters for MCMC estimator are shown in Table~\ref{t:mcmc}. These parameters have similar meanings to those in the \code{stan()} function in the \CRANpkg{rstan} package \citep{rstan}, and detailed explanations of their effects can be found in the Stan reference manual \citep{stan}. By default, \pkg{SPQR} uses \code{algorithm="NUTS"} to approximate the posterior distribution of $\mathcal{W}$ as it adaptively selects both the leap-frog discretization step-size $\epsilon$ as well as the number of steps $L_{\epsilon}$ which are crucial to the sampling efficiency of HMC. The value of step-size is also affected by the target Metropolis acceptance rate \code{delta}. When the NN is large and its posterior geometry is complex, a larger \code{delta} allows NUTS to explore the posterior more carefully using smaller steps. However, a smaller step-size may require a larger \code{max.treedepth} to avoid premature stopping which will significantly increase the algorithm run-time.

Finally, the argument \code{verbose} determines whether training progress should be printed, \code{seed} sets the seed for random number generation when reproducibility is desired, and \code{...} allows any of the control parameters to be specified directly in the function call instead of in \code{control}.

\begin{table}[htbp]
\begin{center}
\caption{\textbf{Control parameters for MCMC}. These parameters are similar to those in \code{stan()} in \pkg{rstan}. Detailed explanations can be found in the Stan reference manual. }\label{t:mcmc}
\setlength\extrarowheight{2pt} % for a bit of visual "breathing space"
\begin{tabular}{p{0.24\textwidth}p{0.7\textwidth}}
\toprule
Parameter & Description \\
\midrule
\code{algorithm}& The sampling algorithm; \code{"HMC"}: Hamiltonian Monte Carlo with dual-averaging, \code{"NUTS"}: No-U-Turn sampler (default).\\
\code{iter}& The number of iterations (including warmup). Default is 2000.\\
\code{warmup}& The number of warmup/burn-in iterations for step-size and mass matrix adaptation. Default is 500.\\
\code{thin}& The period for saving post-warmup samples. The default is 1.\\
\code{stepsize}& The discretization interval/step-size $\epsilon$ of leap-frog integrator. Default is \code{NULL} which means it will be adaptively selected during warmup iterations.\\
\code{metric}& The mass matrix $\bM$; \code{"unit"}: diagonal matrix of ones, \code{"diag"}: diagonal matrix with positive diagonal entries estimated during warmup iterations (default), \code{"dense"}: a dense, symmetric positive definite matrix during warmup iterations.\\
\code{delta}& The target Metropolis acceptance rate. Default is 0.9.\\
\code{max.treedepth}& The maximum tree depth in NUTS. The default is 6.\\
\code{int.time}& The integration time $t$ in HMC. The number of leap-frog steps is then calculated as $L_{\epsilon}=\lfloor t/\epsilon\rfloor$. Default is 1.\\
\bottomrule
\end{tabular}
\end{center}
\end{table}

\subsection{Cross-validation function}

As seen above, the Adam routine used to compute MLE and MAP estimators has numerous potential tuning parameters, such as the learning rate and batch size. In addition, the number of basis functions and hidden neurons will also affect the quality of the estimator. To allow users to select the best model by comparing different model configurations, we provide the \code{cv.SPQR()} function that calculates the cross-validation (CV) error for a given configuration of the MLE or MAP estimator. In addition to all arguments of \code{SPQR}, \code{cv.SPQR()} takes the argument \code{folds} which are pre-computed folds on which CV can be performed. It should be noted that \code{cv.SPQR()} itself does not perform model selection, but can be used as a building block in a grid search loop. There are many ways to generate pre-computed CV folds, such as the \code{createFolds()} function in the \CRANpkg{caret} package \citep{caret}. For users' convenience, we provide the function \code{createFolds.SPQR()} which is equivalent to \code{createFolds()} function in \pkg{caret} but returns an unnamed list. Among all control parameters listed in Table~\ref{t:adam}, we recommend tuning \code{lr}, in addition to \code{n.knots} and \code{n.hidden}. A reasonable range of values to consider is \code{n.knots} $\in\{8,10,12\}$, \code{n.hidden} $\in\{8, 10, 15\}$ and \code{lr} $\in\{e^{-6},e^{-5},e^{-4},e^{-3}\}$.

The \code{cv.SPQR()} function is only applicable to MLE and MAP estimators. For the MCMC estimator, prediction accuracy can be measured by the expected log pointwise
predictive density (ELPD) which can be estimated by Bayesian leave-one-out (LOO) CV  or by widely applicable information criterion \citep[WAIC,][]{vehtari2017}. Calculation of these statistics is integrated in the \code{summary()} functionality and described in the following section. Once convergence is achieved, we recommend running separate chains with different combinations of \code{n.knots} and \code{n.hidden} and select the best model with the highest ELPD.

\subsection{Helper functions for "SPQR" object}

The \code{SPQR()} function returns an object of S3 class \code{"SPQR"}, a compound list that contains the fitted model besides various other information. The \code{summary()} function summarizes the output produced by \code{SPQR()} and structures them in a more organized way to be examined by the user. For \code{SPQR()} fitted using any of the four methods, the output returned by \code{summary()} contains the estimation method, the run-time, and the NN architecture. For MLE and MAP estimators, the output contains the training and validation loss of the final (best if early stopping is used) model, and selected information about the optimizer such as learning rate and batch size. For MAP and MCMC estimators, the output contains the variance hyper-prior model. For MCMC estimator, \code{summary()} calculates various diagnostic statistics that can be used to evaluate the fit of the model and convergence of the MCMC chain. Specifically, we calculate Bayesian LOO-CV and WAIC using the \CRANpkg{loo} package \citep{loo} for model comparison, and the average Metropolis acceptance ratio and the number of divergences of post-warmup iterations to determine if the chain is reliable. The output of \code{summary()} is also an object of S3 class \code{"summary.SPQR"}, to which \code{print.summary()} can be applied to print the aforementioned contents in a user-friendly way. The function \code{print.summary()} has an optional argument \code{showModel} that when set to \code{TRUE}, additionally prints the NN architecture by layer. Instead of using \code{summary} and \code{print.summary}, the user can achieve the same goal by directly \code{print()} the \code{"SPQR"} object in one function call.

The \code{coef()} function outputs the estimated spline coefficients $\theta_k(\bX,\widehat{\mathcal{W}})$ given \code{X}. For MCMC estimator, these will be the posterior means of $\theta_k(\bX,\mathcal{W})$. The \code{predict()} function computes different estimates of the conditional distribution. When \code{type="PDF"} or \code{type="CDF"}, it computes $f(Y|\bX,\widehat{\mathcal{W}})$ or $F(Y|\bX,\widehat{\mathcal{W}})$ for every combination of \code{X} and \code{Y}; when \code{type="QF"}, it computes $Q(\tau|\bX,\hat{\mathcal{W}})$ for every combination of \code{X} and \code{tau}. The argument \code{Y} is optional, and when left unspecified \code{predict()} will use \code{nY} to define a grid on [0,1] for estimation. This is useful when the user wants to estimate the full PDF/CDF/QF curve. For MCMC estimator, two additional arguments are available: \code{ci.level} and \code{getAll}. The argument \code{ci.level} allows the user to extract \code{ci.level}$\times$100\% credible intervals of the estimates in addition to the posterior means, whereas setting \code{getAll=TRUE} allows the user to extract all posterior samples of the corresponding estimates.

\subsection{Quantile accumulated local effect (QALE)}

The function \code{QALE()} is largely based on the \code{ALEPlot()} function in the \CRANpkg{ALEPlot} package \citep{ALEPlot}, but adapted for the quantile regression setting to compute ALEs at different quantiles. The function takes the following arguments,
\begin{example}
  QALE(object, var.index, tau=seq(0.1,0.9,0.1), n.bins=40, ci.level=0, 
    getAll=FALSE, pred.fun=NULL).
\end{example}
The argument \code{object} corresponds to the fitted SPQR object of class \code{"SPQR"}. The \code{var.index} is a numeric scalar or length-two vector of indices of the covariates for which the ALEs will be calculated. When \code{length(var.index) = 1}, the function calculates the main effects for \code{X[,var.index]}; when \code{length(var.index) = 2}, the function calculates interaction effects for \code{X[,var.index[1]]} and \code{X[,var.index[2]]}. The argument \code{tau} is a numeric vector of quantile levels at which the ALEs will be calculated. The argument \code{n.bins} is a numeric scalar that specifies the maximum number of intervals into which the covariate range is divided when calculating the ALEs. The actual number of intervals depends on the number of unique values in \code{X[,var.index]}. When \code{length(var.index) = 2}, \code{n.bins} is applied to both covariates. The arguments \code{ci.level} and \code{getAll} allow uncertainty analysis of the calculated ALEs, but are only implemented for main effects for a single covariate. Finally, the argument \code{pred.fun} accepts a custom quantile prediction function that will be used instead of the built-in \code{predict()} function for calculating ALE. This can be useful when the user wants to compare the QALE calculated using SPQR to that using other quantile regression models, or maybe that using the true model in a simulation study.  

\subsection{Plot functions}

Various plot functions are implemented to visualize results from model prediction, model diagnostics and model interpretation. All of them take directly the \code{"SPQR"} object as input and return a \code{"ggplot"} object that can be further customized using layers from the \CRANpkg{ggplot2} package \citep{ggplot2}. The \code{plotEstimator()} function computes and plots the estimated PDF/CDF/QF curve for a single observation. The \code{plotGOF()} function plots the quantiles of probability integral transform of the observed data against the quantiles of a uniform distribution. Let $U_i=F(Y_i|\bX_i,\widehat{\mathcal{W}})$ be the estimated CDF of the $i$th observation in the dataset. By the probability integral transform, if the observed data indeed distribute according to $F(Y|\bX,\widehat{\mathcal{W}})$ then the sample $U_i,\ i\in\{1,...,n\}$ should correspond to independent samples of a standard uniform distribution, i.e., $U_1,...,U_n\iid\mathcal{U}(0,1)$. Thus the Quantile-Quantile plot (Q-Q plot) created by \code{plotGOF()} can be used for visually checking the goodness-of-fit of the SPQR estimator. When SPQR is fitted with \code{method = "MCMC"}, the function \code{plotMCMCtrace()} can be used to show trace plot of a \code{target} to examine the convergence and autocorrelation of the MCMC chain. Available \code{target} are \code{"loglik"} for the log-likelihood, and \code{"PDF"}, \code{"CDF"} and \code{"QF"} for corresponding estimate for a single observation. The function also takes an optional argument \code{window} which allows examination of different parts of the chain. The functions \code{plotQALE()} and \code{plotQVI()} support the \code{QALE()} function in helping the user understand how the covariates affect different quantiles, and thus accept similar arguments. In particular, the function \code{plotQALE()} plots the estimated quantile ALE main effects when \code{length(var.index) = 1} and quantile ALE interaction effects when \code{length(var.index) = 2}, whereas the function \code{plotQVI()} compares the quantile ALE-induced variable importance (VI) between all considered covariates. The \code{var.index} argument can be left unspecified in \code{plotQVI()} in which case all covariates in the data set are considered. Similar to the \code{predict()} function, most plot functions accept arguments \code{ci.level} or \code{getAll} (or both) for uncertainty quantification when SPQR is fitted with \code{method = "MCMC"}. Table~\ref{t:plot} summarizes the implementation of uncertainty quantification for plot functions in \pkg{SPQR}. Finally, the function \code{autoplot()} provides a wrapper for all the plot functions mentioned above.

\begin{table}[htbp]
\begin{center}
\caption{Implementation of uncertainty quantification for plot functions in \pkg{SPQR}.}\label{t:plot}
\begin{tabular}{lcc}
\toprule
Name of the function & \code{ci.level} & \code{getAll} \\
\midrule
\code{plotEstimator()} & $\checkmark$ & $\checkmark$ \\
\code{plotGOF()} & $\times$ & $\checkmark$\\
\code{plotMCMCtrace()} & N/A & N/A\\
\code{plotQALE()} & $\checkmark$ & $\checkmark$\\
\code{plotQVI()} & $\checkmark$ & $\times$\\
\bottomrule
\end{tabular}
\end{center}
\end{table}

\section{Examples}
\subsection{Simulation}
\begin{figure}[htbp]
  \centering
  \includegraphics[width=0.8\linewidth]{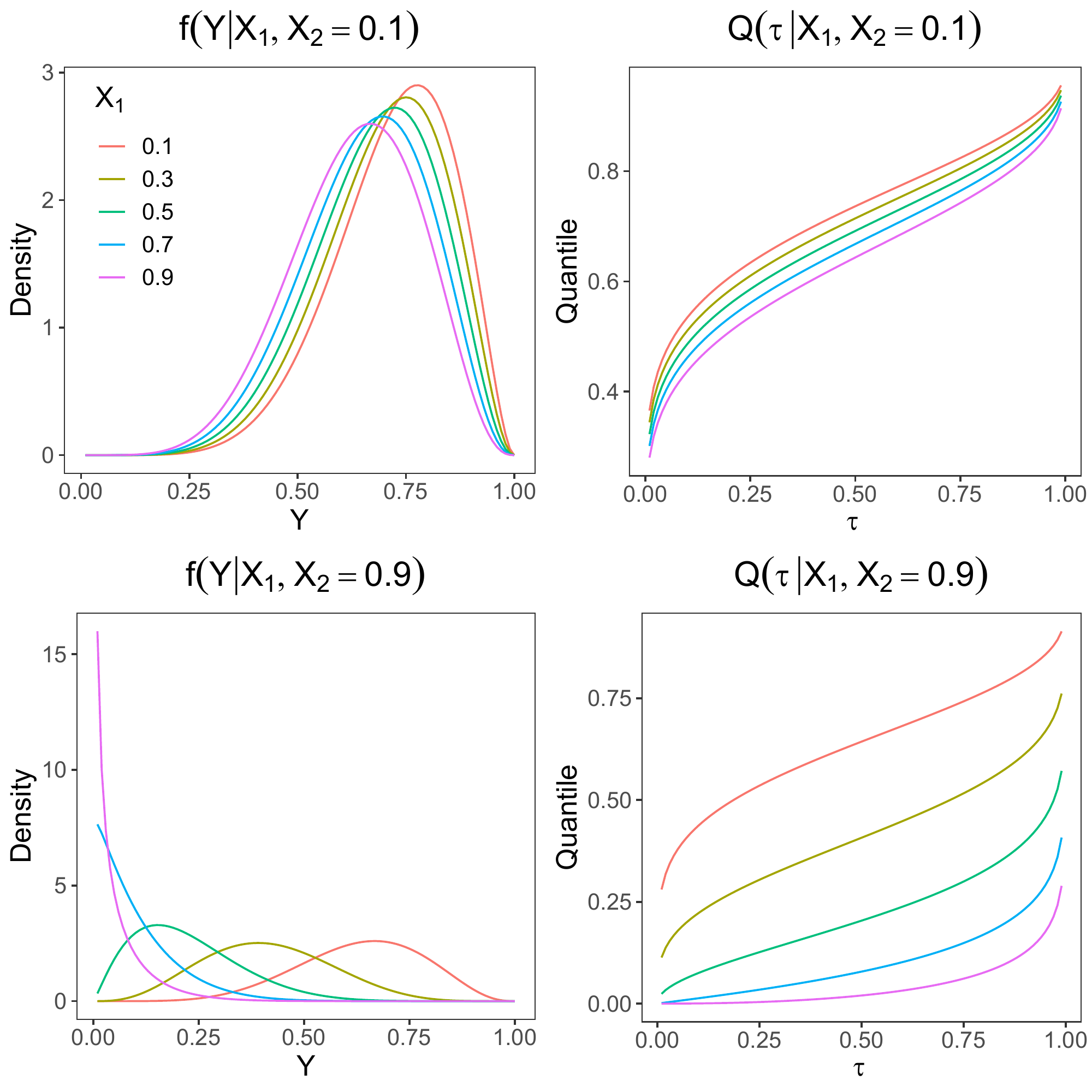}
  \caption{True conditional density and quantile functions of \eqref{ex1} for different combinations of $X_1$ and $X_2$.}
  \label{f:sim1dqf}
\end{figure}

We start with a simple simulation study to demonstrate the effectiveness of SPQR in estimating conditional density and quantile functions, as well as usages of functions introduced in the previous section. We consider a three dimensional covariate $\bX = (X_1,X_2,X_3)$ with variables independently generated from a uniform distribution. The response $Y$ follows a Beta distribution whose shape parameters are functions of $X_1$ and $X_2$, the third covariate is irrelevant.
\begin{equation}\label{ex1}
    \begin{split}
        X_j&\iid\mathcal{U}(0,1), \ j=1,...,3\\
        Y &\sim \mathcal{B}eta\lpth10\mbox{expit}\lcbk 1-5X_1X_2\rcbk,10\lsbk1-\mbox{expit}\lcbk 1-5X_1X_2\rcbk\rsbk\rpth.
    \end{split}
\end{equation}
Here $\mbox{expit}(u)=1/(1+e^{-u})$ is the inverse logistic link function. Based on the definition of Beta distribution, the following properties of the conditional distribution $f(Y|\bX)$ can be obtained
\begin{equation}
    \begin{split}
        \mathbb{E}(Y|\bX) &\propto \mbox{expit}\lcbk 1-5X_1X_2\rcbk\\
        \mbox{Var}(Y|\bX) &\propto \mbox{expit}\lcbk 1-5X_1X_2\rcbk\lsbk1-\mbox{expit}\lcbk 1-5X_1X_2\rcbk\rsbk\\
        \mbox{Skewness}(Y|\bX) &\propto -\frac{\sqrt{\mbox{expit}\lcbk 1-5X_1X_2\rcbk}}{\sqrt{1-\mbox{expit}\lcbk 1-5X_1X_2\rcbk}}.
    \end{split}
\end{equation}
That is, the location, scale and shape of $f(Y|\bX)$ all depend on $\bX$. Figure~\ref{f:sim1dqf} shows the true density and quantile functios of $Y$ as conditioned on different combinations of $X_1$ and $X_2$. We can clearly observe a varying effect of $\bX$ on the density and quantile functions of $Y$.

\begin{figure}[htbp]
  \centering
  \includegraphics[width=\linewidth]{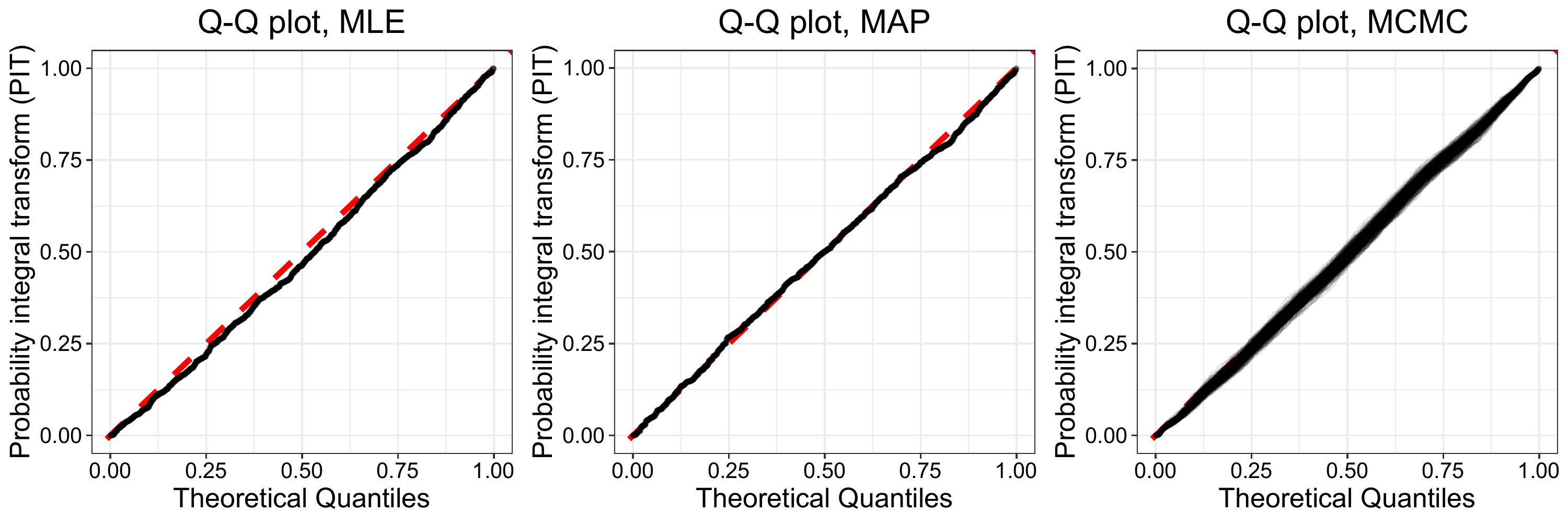}
  \caption{Examples of plot produced by \code{plotGOF()}. Goodness-of-fit test for SPQR estimators.}
  \label{f:sim1qq}
\end{figure}

\begin{figure}[htbp]
  \centering
  \includegraphics[width=\linewidth]{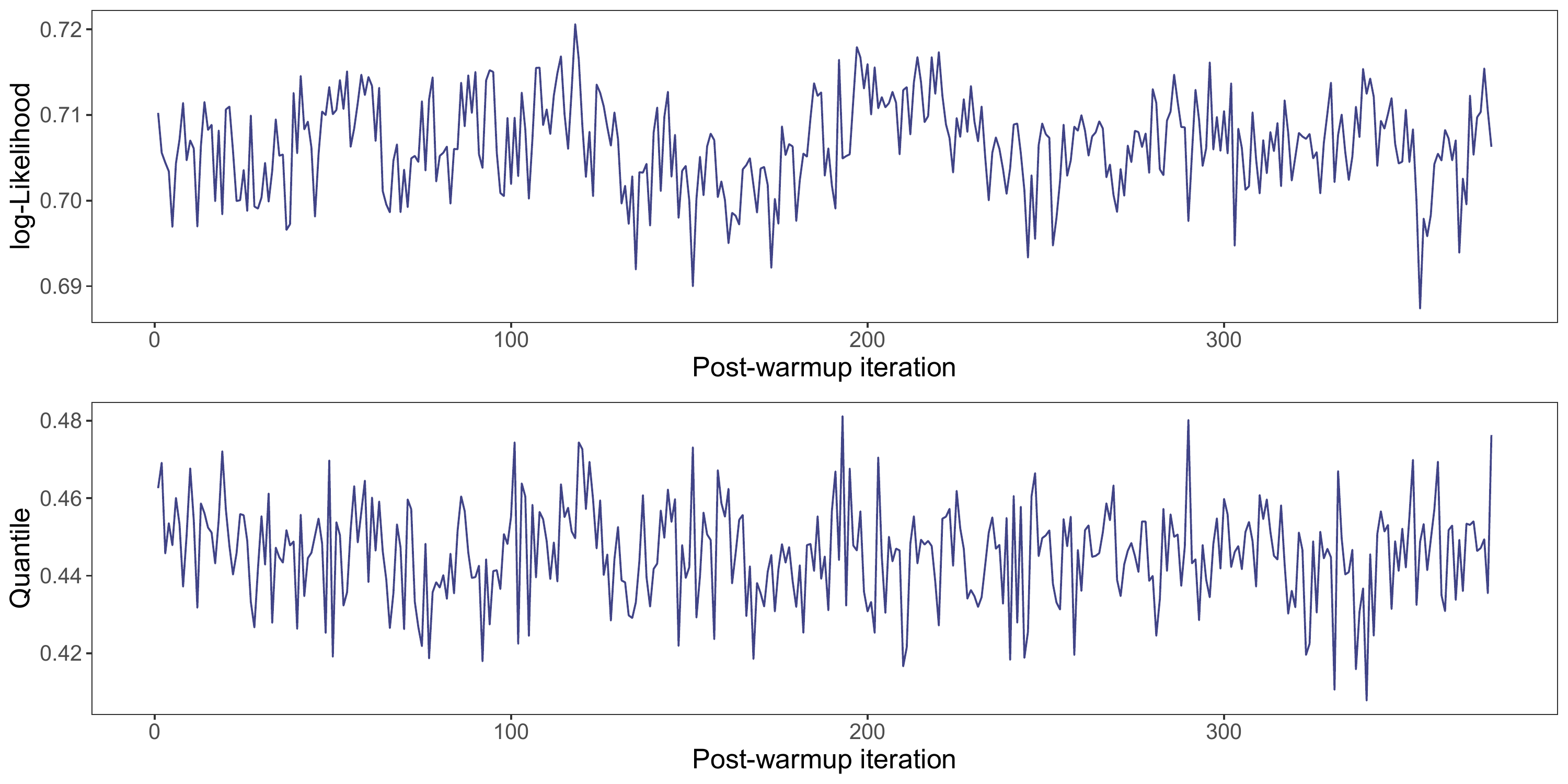}
  \caption{Examples of traceplot produced by \code{plotMCMCtrace()}.}
  \label{f:sim1trace}
\end{figure}

We generate $n=1000$ samples from \eqref{ex1} and fit SPQR using the MLE, MAP, and MCMC methods. For all three methods, the default configuration of \code{n.knots=10}, \code{n.hidden=c(10)} and \code{activation="tanh"} are used. For the MLE and MAP methods, the learning rate is selected from $\{e^{-6},e^{-5},e^{-4},e^{-3}\}$ using 5-fold cross-validation, and the gradients are calculated using mini-batches of size 256 with a maximum training time of 500 epochs. For the MAP method, the default \code{prior="ARD"} is used.
\begin{example}
    ## Generate simulated data
    set.seed(919)
    n     <- 1000
    p     <- 3
    X     <- matrix(runif(n*p),n,p)
    expit <- 1/(1+exp(-1+5*X[,1]*X[,2]))
    Y     <- rbeta(n,10*expit,10*(1-expit))
    ## Specifying MLE control parameters
    mle.control <- list(batch.size = 256, epochs = 500, use.GPU = TRUE)
    ## Generate CV folds
    folds <- createFolds.SPQR(Y, nfold=5)
    ## Select optimal learning rate using 5-fold CV
    lr.grid <- exp(-6:-3)
    cve <- sapply(lr.grid, FUN=function(lr) {
        mle.control$lr <- lr
        cv.out <- cv.SPQR(X=X, Y=Y, folds=folds, method="MLE", control=mle.control)
        cv.out$cve
    })
    mle.control$lr <- lr.grid[which.min(cve)]
    mle.fit <- SPQR(X=X, Y=Y, method="MLE", control=mle.control)
    
    ## Specifying MAP control parameters
    map.params <- mle.params
    cve <- sapply(lr.grid, FUN=function(lr) {
        map.control$lr <- lr
        cv.out <- cv.SPQR(X=X, Y=Y, folds=folds, method="MAP", control=map.control)
        cv.out$cve
    })
    map.control$lr <- lr.grid[which.min(cve)]
    map.fit <- SPQR(X=X, Y=Y, method="MAP", control=map.control)
\end{example}
For the MCMC method, the posterior distribution is approximated using NUTS. We run NUTS for a total of 1000 iterations, discard the first 250 as warm-ups, and save every other iteration. 
\begin{example}
  ## Specifying MCMC control parameters
  mcmc.control <- list(iter = 1000, warmup = 250, thin = 2)
  mcmc.fit     <- SPQR(X=X, Y=Y, method="MCMC", control=mcmc.control)
\end{example}
The function \code{print()} returns a short summary of the results of the fitted object. The summary depends on the fitted method. Here we show summary for models fitted with \code{method="MLE"} and \code{method="MCMC"}. The summary for \code{method="MAP"} is mostly the same as that for \code{method="MLE"} and thus omitted.
\begin{example}
mle.fit

#> SPQR fitted using MLE approach
#> 
#> Learning rate: 0.04978707
#> Batch size: 256
#> 
#> Loss:
#>   train = -140.2645,  validation = -142.1546
#> 
#> Elapsed time: 0.22 minutes

mcmc.fit

#> SPQR fitted using MCMC approach with ARD prior
#> 
#> MCMC diagnostics:
#>   Final acceptance ratio is 0.91 and target is 0.9
#> 
#> Expected log pointwise predictive density (elpd) estimates:
#>   elpd.LOO = 683.4493,  elpd.WAIC = 683.9026
#> 
#> Elapsed time: 3.05 minutes
\end{example}
An overview of the NN architecture can be additionally printed by setting the argument \code{showModel} to \code{TRUE}.
\begin{example}
print(mle.fit, showModel = TRUE)

#> SPQR fitted using MLE approach
#> 
#> Learning rate: 0.04978707
#> Batch size: 256
#> 
#> Model specification:
#>     Layers
#>    Input Output Activation
#>        3     10       tanh
#>       10     10    softmax
#> 
#> Loss:
#>   train = -140.2645,  validation = -142.1546
#> 
#> Elapsed time: 0.22 minutes
\end{example}
The function \code{plotGOF()} uses a Q-Q plot to inspect the alignment of the probability integral transform (PIT) and the uniform distribution, which can be used to visually check the goodness of fit of \code{SPQR()}. Figure~\ref{f:sim1qq} shows the goodness of fit of the three SPQR estimators. For SPQR fitted with \code{method="MCMC"}, the argument \code{getAll} is used to additionally plot all posterior samples of the Q-Q plot. The plots show that the distribution of PIT is most similar to the uniform distribution for the MCMC estimator, suggesting that the model based on the MCMC method best fits the observed data.
\begin{example}
  ## Q-Q plot of PIT for goodness of fit test
  plotGOF(mle.fit)  + ggtitle("Q-Q plot, MLE")
  plotGOF(map.fit)  + ggtitle("Q-Q plot, MAP")
  plotGOF(mcmc.fit, getAll = TRUE) + ggtitle("Q-Q plot, MCMC")
\end{example}
The reliability of the MCMC estimator can be further assessed by the function \code{plotMCMCtrace()}, which plots traceplots that can be used to visually inspect the sampling behavior and convergence of the post-warmup chain. The code below provides two examples of such usage, one uses \code{target="loglik"} to plot the traceplot of the log-likehood; the other uses \code{target="QF"}, \code{X=c(0.5,0.5,0.5)} and \code{tau=0.5} to plot the traceplot of the estimated median when $\bX=(0.5,0.5,0.5)^\intercal$. The results are shown in Figure~\ref{f:sim1trace} and provide visual evidence that the chain converged.
\begin{example}
  ## Traceplot of log-likelihood
  plotMCMCtrace(mcmc.fit, target = "loglik")
  ## Traceplot of estimated quantile
  plotMCMCtrace(mcmc.fit, target = "QF", X = c(0.5,0.5,0.5), tau = 0.5)
\end{example}

\begin{figure}[htbp]
  \centering
  \includegraphics[width=\linewidth]{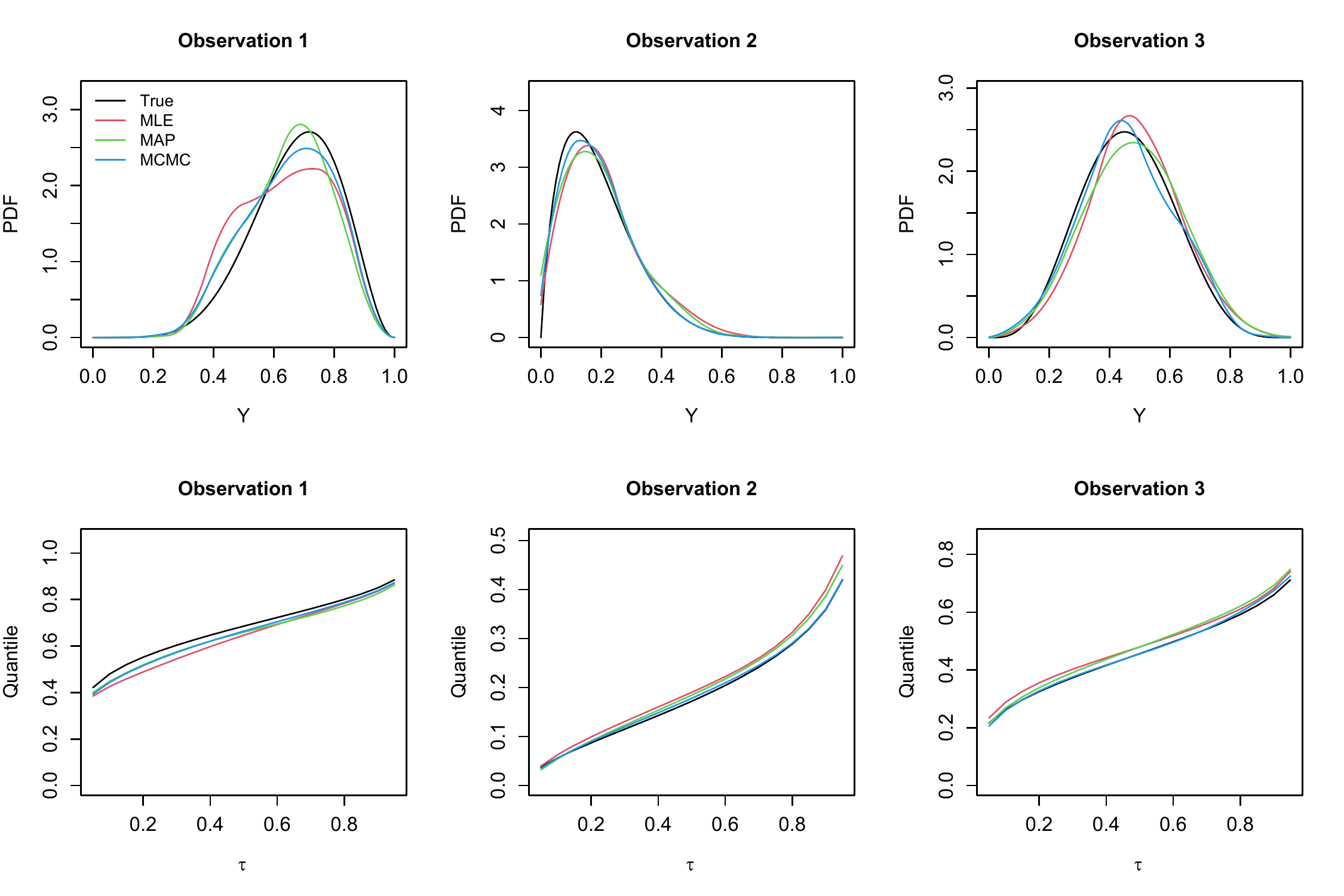}
  \caption{Estimated and true PDFs (top) and QFs (bottom) for 3 out-of-sample observations.}
  \label{f:sim1dqfe}
\end{figure}

\begin{figure}[htbp]
  \centering
  \includegraphics[width=\linewidth]{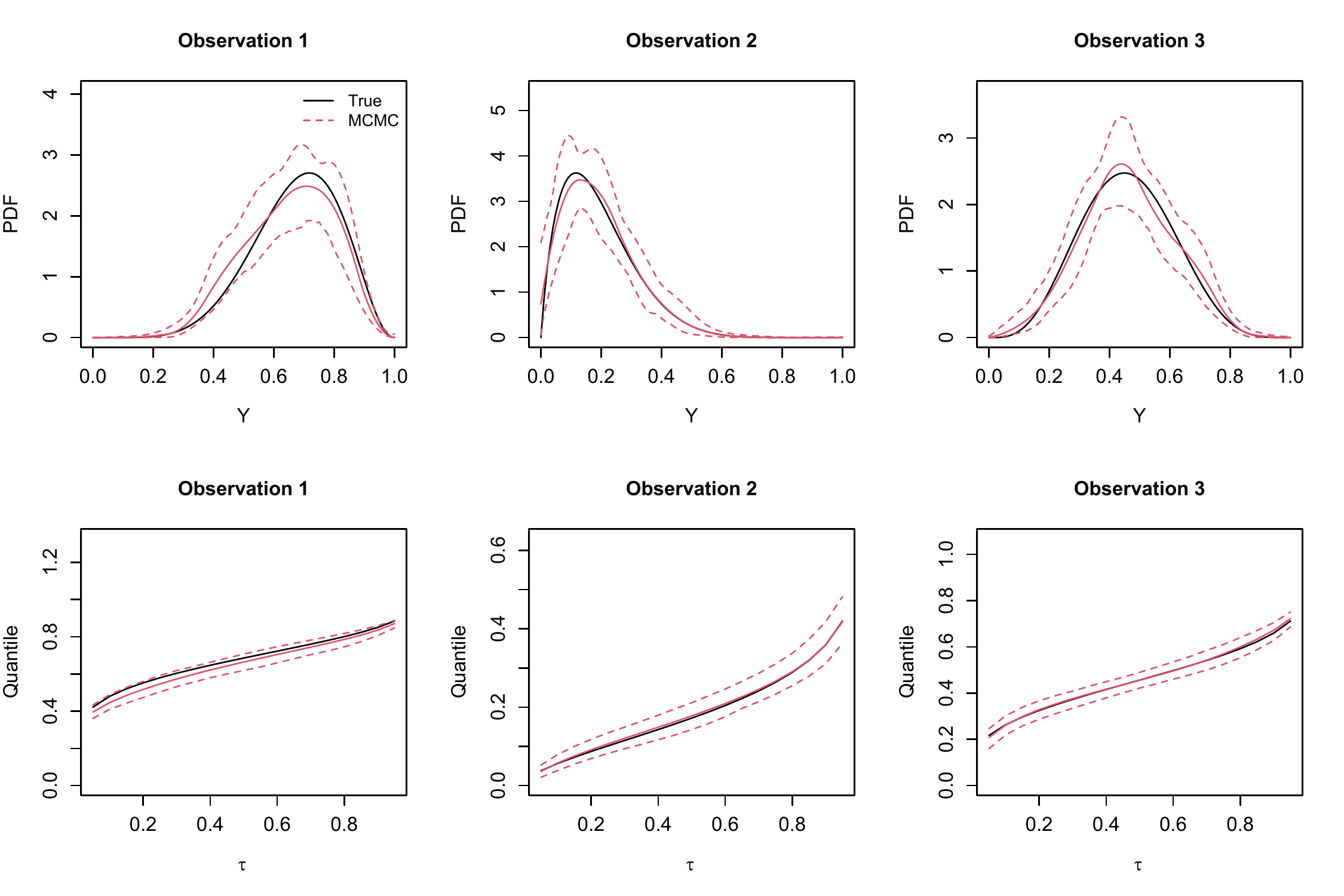}
  \caption{Estimated and true PDFs (top) and QFs (bottom), along with 95\% credible bands, for 3 out-of-sample observations.}
  \label{f:sim1dqfci}
\end{figure}

\begin{figure}[htbp]
  \centering
  \includegraphics[width=0.9\linewidth]{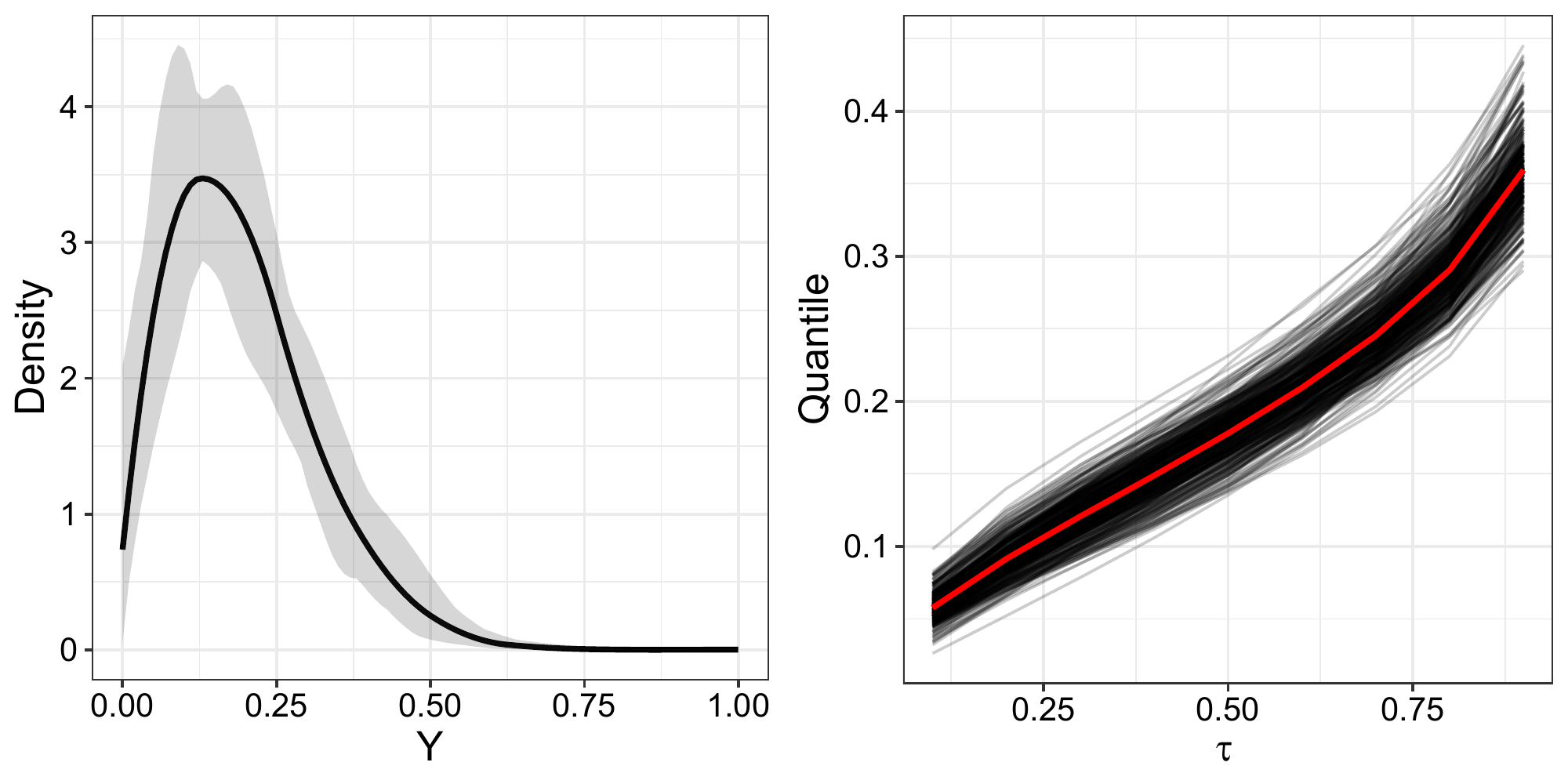}
  \caption{Examples of plot produced by \code{plotEstimator()}. Left: estimated PDF along with 95\% credible bands. Right: estimated QF along with all posterior samples.}
  \label{f:sim1auto}
\end{figure}

To investigate the numerical performance of the fitted SPQR models, we use the function \code{predict()} to compute the predicted conditional PDFs and QFs for 3 out-of-sample observations and compare them with their true values.
\begin{example}
  ## Prediction based on 3 out-of-sample observations
  set.seed(3)
  X_test <- matrix(runif(3*3), nrow = 3)
  ## PDFs
  ## Not specifying `Y` means to estimate on a fine grid
  ## Default is Y <- seq(0,1,length=101)
  pdf.mle  <- predict(mle.fit, X = X_test, type = "PDF")
  pdf.map  <- predict(map.fit, X = X_test, type = "PDF")
  pdf.mcmc <- predict(mcmc.fit, X = X_test, type = "PDF")
  ## QFs
  qf.mle  <- predict(mle.fit, X = X_test, type = "QF")
  qf.map  <- predict(map.fit, X = X_test, type = "QF")
  qf.mcmc <- predict(mcmc.fit, X = X_test, type = "QF")
  
  ## Function to compute true PDFs
  dYgivenX <- function(y, X){
    expit <- 1/(1+exp(-1+5*X[1]*X[2]))
    out   <- dbeta(y, 10*expit, 10*(1-expit))
    return(out)}
  ## Compare the predicted PDFs with their true values
  yyy <- seq(0, 1, length.out = 101)
  par(mfrow = c(2,3))
  for(i in 1:3){
    pdf0 <- dYgivenX(yyy, X_test[i,]) # True
    pdf1 <- pdf.mle[i,]     # MLE
    pdf2 <- pdf.map[i,]     # MAP
    pdf3 <- pdf.mcmc[i,]    # MCMC
    plot(yyy, pdf0, ylim=c(0,1.2*max(pdf0)), type="l",
         xlab="Y", ylab="PDF", main=paste("Observation",i),
         cex.lab=1.2, cex.axis=1.2)
    lines(yyy, pdf1, col=2)
    lines(yyy, pdf2, col=3)
    lines(yyy, pdf3, col=4)
    if(i == 1) {
      legend("topleft", c("True","MLE","MAP","MCMC"), lty=1, col=1:4, bty="n")
    }}

  ## Function to compute true QFs
  qYgivenX <- function(tau,X){
    expit <- 1/(1+exp(-1+5*X[1]*X[2]))
    out   <- qbeta(tau, 10*expit, 10*(1-expit))
    return(out)}
  ## Compare the predicted QFs with their true values
  for(i in 1:3){
    qf0 <- qYgivenX(tau,X_test[i,]) # True
    qf1 <- qf.mle[i,]     # MLE
    qf2 <- qf.map[i,]     # MAP
    qf3 <- qf.mcmc[i,]    # MCMC
    plot(tau,qf0,ylim=c(0,1.2*max(qf0)),type="l",
         xlab="tau",ylab="Quantile",main=paste("Observation",i),
         cex.lab=1.2, cex.axis=1.2)
    lines(tau,qf1,col=2)
    lines(tau,qf2,col=3)
    lines(tau,qf3,col=4)
  }
\end{example}
The estimated and true PDFs are shown in the top row of Figure~\ref{f:sim1dqfe}, and the estimated and true QFs are shown in the bottom row. 
For the MCMC estimator, we also compute the 95\% pointwise credible intervals using the argument \code{ci.level=0.95} and plot them in Figure~\ref{f:sim1dqfci}. These plots show that SPQR successfully captured the varying effect of $\bX$ on the distribution of $Y$. 
\begin{example}
  ## PDF point estimates and credible intervals
  pdf.mcmc <- predict(mcmc.fit, X_test, type="PDF", ci.level=0.95)
  ## QF point estimates and credible intervals
  qf.mcmc <- predict(mcmc.fit, X_test, type="QF", ci.level=0.95)
  
  ## Compare the predicted PDFs with their true values
  par(mfrow=c(2,3))
  for (i in 1:3) {
    pdf0    <- dYgivenX(yyy,X_test[i,]) # True
    pdf1.lb <- pdf.mcmc[i,,"lower.bound"]
    pdf1    <- pdf.mcmc[i,,"mean"]
    pdf1.ub <- pdf.mcmc[i,,"upper.bound"]
    plot(yyy,pdf0,ylim=c(0,1.5*max(pdf0)),type="l",
         xlab="Y",ylab="PDF",main=paste("Observation",i),
         cex.lab=1.2, cex.axis=1.2)
    lines(yyy,pdf1,col=2)
    lines(yyy,pdf1.lb,col=2,lty=2)
    lines(yyy,pdf1.ub,col=2,lty=2)
    if (i==1) {
      legend("topright",c("True","MCMC"),lty=1:2,col=1:2,bty="n")
    }
  }
  ## Compare the predicted QFs with their true values
  for (i in 1:3) {
    qf0    <- qYgivenX(tau,X_test[i,]) # True
    qf1.lb <- qf.mcmc[i,,"lower.bound"]
    qf1    <- qf.mcmc[i,,"mean"]
    qf1.ub <- qf.mcmc[i,,"upper.bound"]
    plot(tau,qf0,ylim=c(0,1.5*max(qf0)),type="l",
         xlab=expression(tau),ylab="Quantile",main=paste("Observation",i),
         cex.lab=1.2, cex.axis=1.2)
    lines(tau,qf1,col=2)
    lines(tau,qf1.lb,col=2,lty=2)
    lines(tau,qf1.ub,col=2,lty=2)
  }
\end{example}
Figures~\ref{f:sim1dqfe} and \ref{f:sim1dqfci} can be easily reproduced using \code{plotEstimator()} which inherits most arguments of \code{predict()}. The codes below show examples of plotting the predicted PDF along with 95\% credible bands and plotting the predicted QF along with all posterior samples. The results are shown in Figure~\ref{f:sim1auto}.
\begin{example}
  ## PDF with 95% credible bands
  plotEstimator(mcmc.fit, X = X_test[2,], type = "PDF", ci.level = 0.95) 
  ## QF with all posterior samples
  plotEstimator(mcmc.fit, X = X_test[2,], type = "QF", getAll = TRUE) 
\end{example}

\begin{figure}[htbp]
  \centering
  \includegraphics[width=\linewidth]{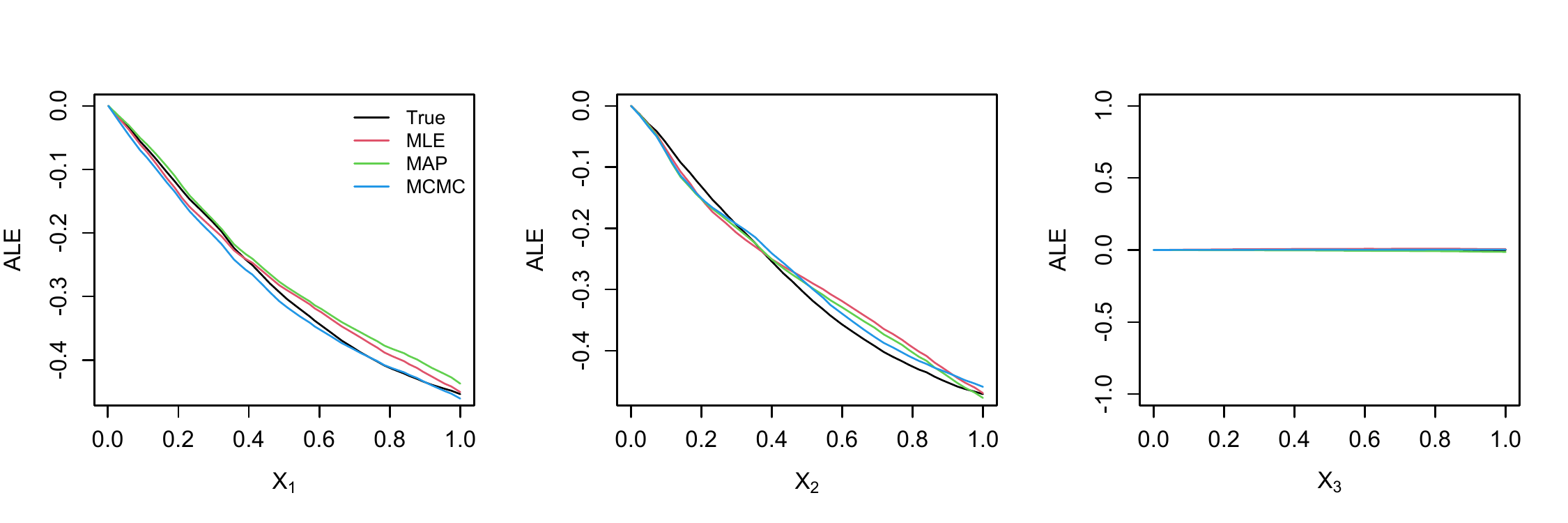}
  \caption{Quantile accumulative local effects (ALEs) for $X_1$, $X_2$ and $X_3$ respectively at $\tau=0.5$.}
  \label{f:sim1ale}
\end{figure}

In most applications where quantile regression is used, understanding the covariate effect on the conditonal quantile/distribution is of paramount importance. The function \code{QALE()} can be used to quantify either the main effect of a single covariate or the interaction effect between two covariates on the predicted quantiles. the codes below show examples of computing the QALE for $X_1$, $X_2$, $X_3$ respectively at $\tau=0.25$ using fitted \code{"SPQR"} objects, as well as an example using custom prediction function (in this case it is the true quantile function). To compute the QALE using a custom prediction function, the user should first define a function that takes \code{X}, the covariate matrix, and \code{tau}, a vector of quantile levels, as inputs and returns an \code{nrow(X)} by \code{length(tau)} matrix of predicted quantiles. The user should then pass the defined function to the \code{pred.fun} argument and \code{list(X=X)} to the \code{object} argument. The results are shown in Figure~\ref{f:sim1ale} and provide visual evidence that SPQR accurately captures the quantile covariate effects.
\begin{example}
  ## Quantile of interest
  tau <- 0.25
  ## A custom prediction function that computes the true quantiles
  pred.fun <- function(X, tau) {
    out <- matrix(nrow=nrow(X), ncol=length(tau))
    for (i in 1:nrow(X)) {
      out[i,] <- qYgivenX(tau, X[i,])
    }
    return(out)
  }
  ## Main effect for X1, X2, X3
  par(mfrow=c(1,3))
  for (j in 1:3) {
    ale.mle  <- QALE(mle.fit, var.index=j, tau=tau)    # MLE
    ale.map  <- QALE(map.fit, var.index=j, tau=tau)    # MAP
    ale.mcmc <- QALE(mcmc.fit, var.index=j, tau=tau)  # MCMC
    ## The following line will work as long as `object` is a list that contains `X`
    ale.ans <- QALE(list(X=X), var.index=j, tau=tau, pred.fun=pred.fun) # True
    plot(ale.ans$x, ale.ans$ALE, type="l", xlab=parse(text=paste0("X[",j,"]")), 
         ylab="ALE", cex.lab=1.2, cex.axis=1.2)
    lines(ale.mle$x, ale.mle$ALE, col=2)
    lines(ale.map$x, ale.map$ALE, col=3)
    lines(ale.mcmc$x, ale.mcmc$ALE, col=4)
    if (j == 1) {
      legend("topright", c("True","MLE","MAP","MCMC"), lty=1, col=1:4, bty="n")
    }
  }
\end{example}

\begin{figure}[htbp]
  \centering
  \includegraphics[width=\linewidth]{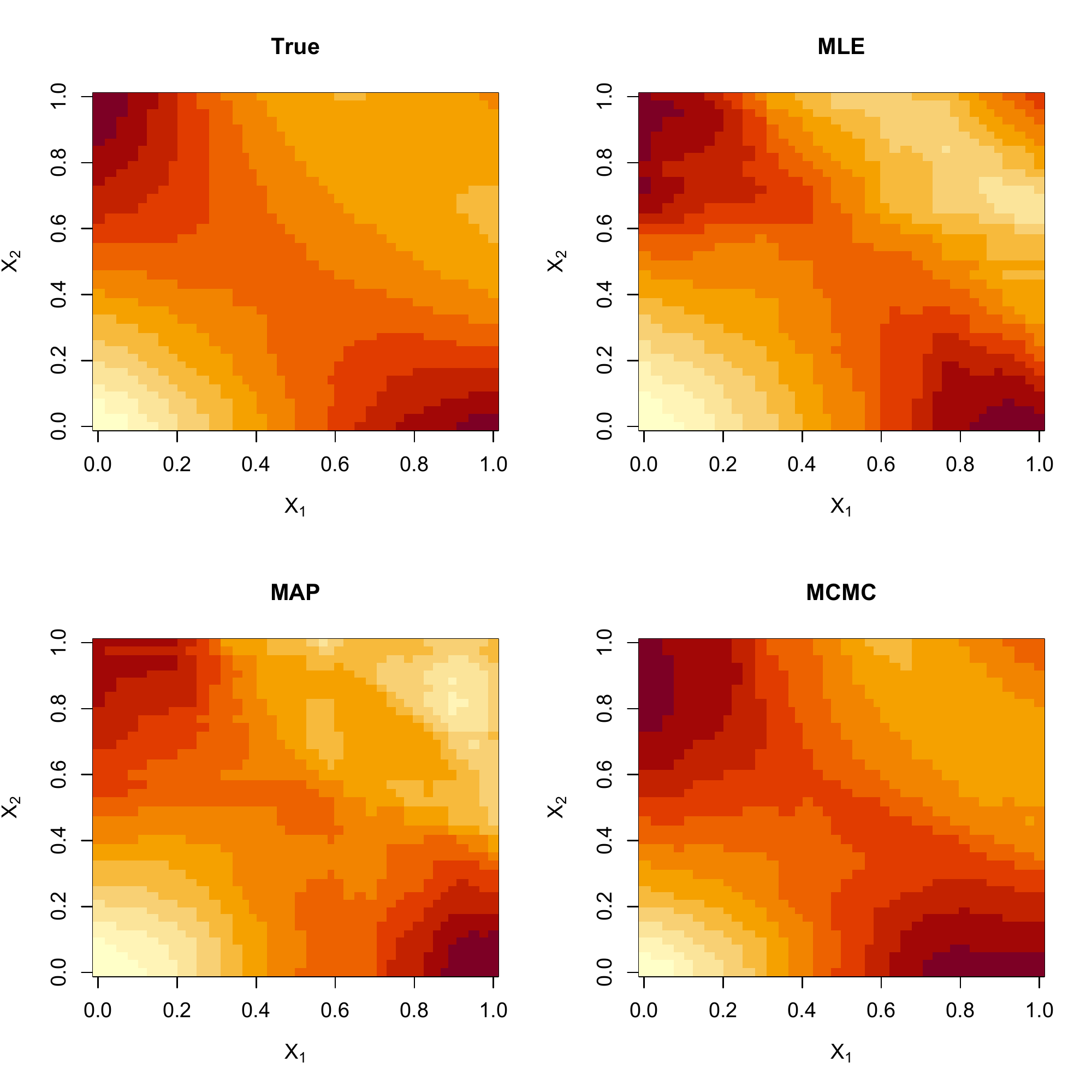}
  \caption{Quantile accumulative local effects (ALEs) interaction effect between $X_1$ and $X_2$ at $\tau=0.5$.}
  \label{f:sim1aleint}
\end{figure}

We also compute the QALE interaction effect between $X_1$ and $X_2$ at $\tau=0.5$. The results are shown in Figure~\ref{f:sim1aleint}. The plots show that SPQR is able to recover the complex interaction effect between $X_1$ and $X_2$.
\begin{example}
  par(mfrow=c(2,2))
  ## Interaction effect between X1 and X2
  ale.mle  <- QALE(mle.fit, var.index=c(1,2), tau=0.5)
  ale.map  <- QALE(map.fit, var.index=c(1,2), tau=0.5)
  ale.mcmc <- QALE(mcmc.fit, var.index=c(1,2), tau=0.5)
  ale.ans  <- QALE(list(X=X), var.index=c(1,2), tau=0.5, pred.fun=pred.fun)
  ## Contour plots for visualizing 2D effects
  image(ale.ans$x[[1]], ale.ans$x[[2]], ale.ans$ALE[,,1], xlab = parse(text="X[1]"),
        ylab = parse(text="X[2]"), main = "True", cex.lab=1.1, cex.axis=1.1)
  image(ale.mle$x[[1]], ale.mle$x[[2]], ale.mle$ALE[,,1], xlab = parse(text="X[1]"),
        ylab = parse(text="X[2]"), main = "MLE", cex.lab=1.1, cex.axis=1.1)
  image(ale.map$x[[1]], ale.map$x[[2]], ale.map$ALE[,,1], xlab = parse(text="X[1]"),
        ylab = parse(text="X[2]"), main = "MAP", cex.lab=1.1, cex.axis=1.1)
  image(ale.mcmc$x[[1]], ale.mcmc$x[[2]], ale.mcmc$ALE[,,1], xlab = parse(text="X[1]"),
        ylab = parse(text="X[2]"), main = "MCMC", cex.lab=1.1, cex.axis=1.1)
\end{example}

\begin{figure}[htbp]
  \centering
  \includegraphics[width=\linewidth]{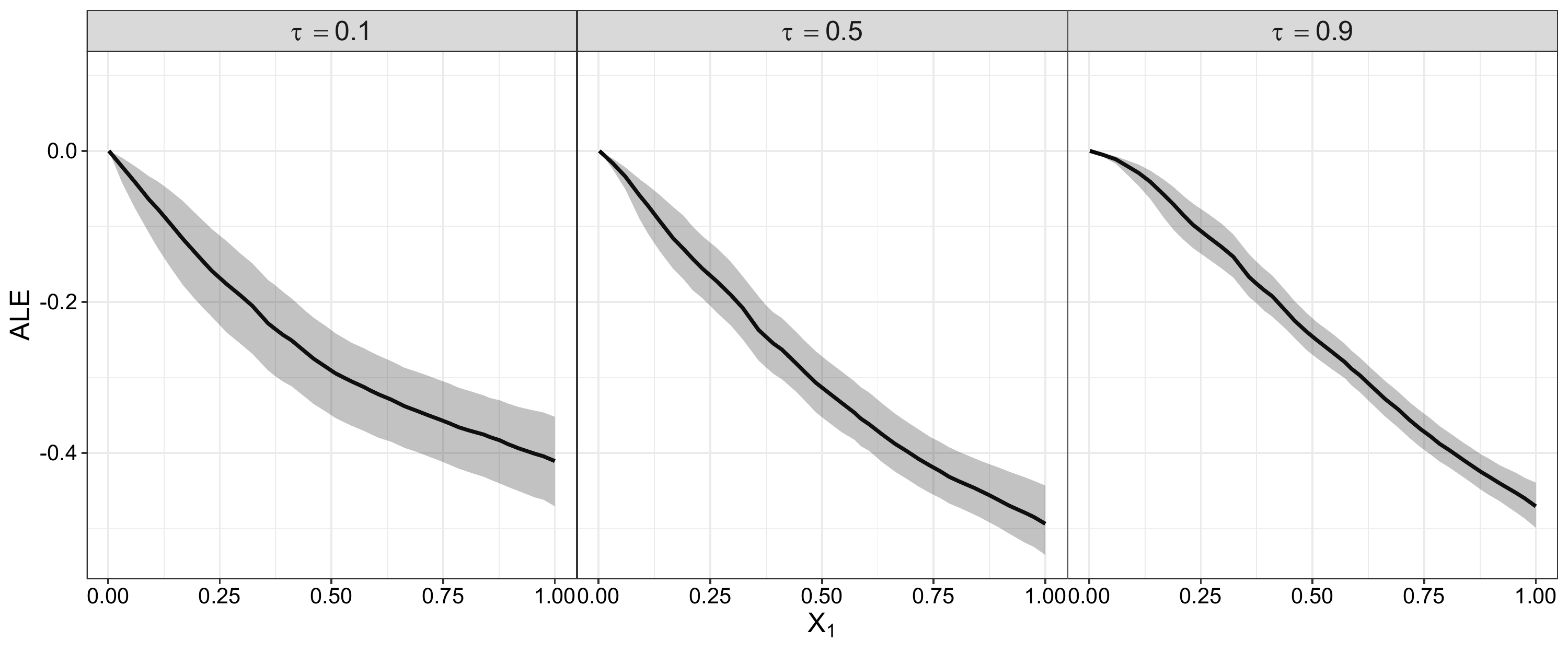}
  \caption{Example of plot produced by \code{plotQALE()}. QALE main effect for $X_1$ at $\tau\in\{0.1,0.5,0.9\}$.}
  \label{f:sim1alepm}
\end{figure}

\begin{figure}[htbp]
  \centering
  \includegraphics[width=\linewidth]{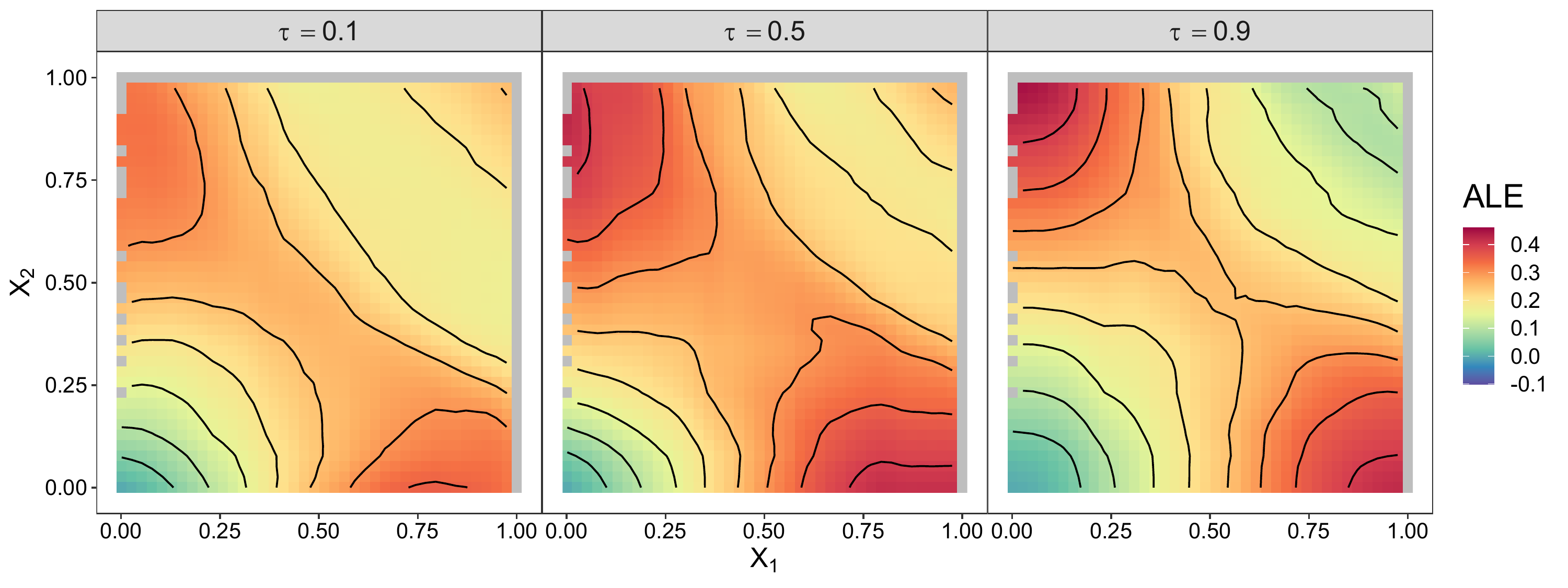}
  \caption{Example of plot produced by \code{plotQALE()}. QALE interaction effect between $X_1$ and $X_2$ at $\tau\in\{0.1,0.5,0.9\}$.}
  \label{f:sim1alepint}
\end{figure}

Figures~\ref{f:sim1ale} and \ref{f:sim1aleint} can be reproduced using the function \code{plotQALE()}. the codes below show two examples. The first one plots the QALE main effect of $X_1$ at $\tau\in\{0.1,0.5,0.9\}$, along with 95\% credible bands. The second one plots the QALE interaction effect between $X_1$ and $X_2$ at $\tau\in\{0.1,0.5,0.9\}$. The results are shown in Figures~\ref{f:sim1alepm} and \ref{f:sim1alepint}, respectively. The results suggest that the main effect of $X_1$ and the interaction effect between $X_1$ and $X_2$ are more prominent on upper quantiles than on lower quantiles.
\begin{example}
  ## Main effect for X1
  plotQALE(mcmc.fit, var.index = 1, tau = c(0.1,0.5,0.9), ci.level = 0.95)
  ## Interaction effect between X1 and X2
  plotQALE(mcmc.fit, var.index = c(1,2), tau = c(0.1,0.5,0.9))
\end{example}

\begin{figure}[htbp]
  \centering
  \includegraphics[width=\linewidth]{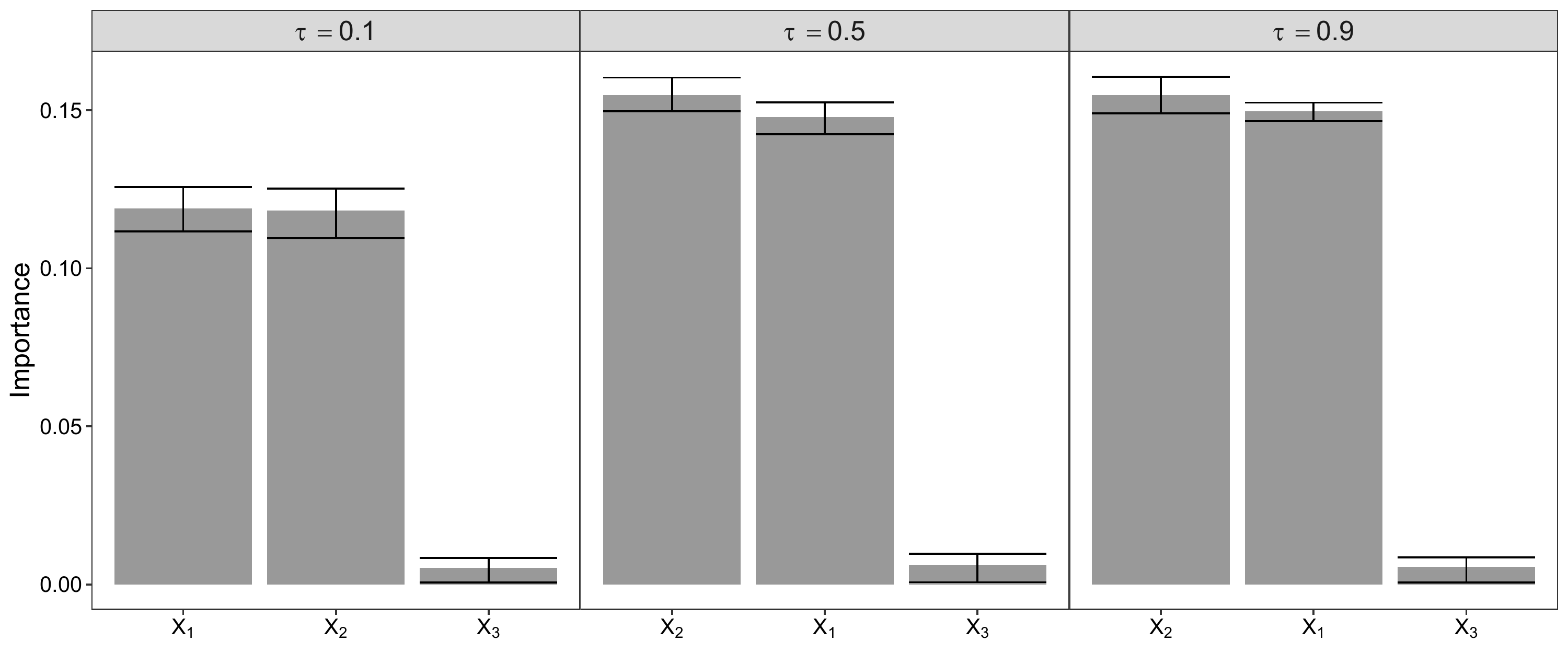}
  \caption{Example of plot produced by \code{plotQVI()}. Quantile variable importance (VI) at $\tau\in\{0.1,0.5,0.9\}$.}
  \label{f:sim1alevi}
\end{figure}

The function \code{plotQVI()} computes the QALE-induced variable importance (VI) measures of each covariate. The covariates are then ranked accordingly in a barplot at each quantile of interest. the codes below show an example of computing the quantile VI for $X_1$, $X_2$ amd $X_3$ at $\tau\in\{0.1,0.5,0.9\}$, respectively. The argument \code{ci.level = 0.95} is used to plot 95\% error bar for each VI measure. The result is shown in Figure~\ref{f:sim1alevi}. The result suggests that $X_1$ and $X_2$ have similar and significant effect on quantiles of $Y$ whereas $X_3$ has no effect. The effects of $X_1$ and $X_2$ are also more prominent on upper quantiles than on lower quantiles. These observations match the truth.
\begin{example}
  ## Quantile variable importance with error bars
  plotQVI(mcmc.fit, tau=c(0.1,0.5,0.9), ci.level=0.95)
\end{example}
\begin{example}
  
\end{example}
\subsection{Australia electricity demand data}

For real data application we consider the electricity demand data from Sydney, Australia. The dataset contains electric energy consumption, recorded from smart meters, of 247 anonymized residential customers during the period between July 3rd 2010 and June 30th 2011. It is available from the R package \pkg{qgam} and has been analyzed by \citet{qgam} in the context of quantile regression. The data can be loaded using the following codes,
\begin{example}
  ## Load the electricity demand data
  data("AUDem", package = "qgam")
  meanDem <- AUDem$meanDem
  head(meanDem)
  
       doy  tod       dem dow     temp                date     dem48
  #> 1 184 18.0 0.8248777 Sat 3.357407 2010-07-03 17:30:00 0.8978636
  #> 2 184 18.5 0.8686110 Sat 2.517073 2010-07-03 18:00:00 0.9417633
  #> 3 184 19.0 0.8519471 Sat 1.898399 2010-07-03 18:30:00 0.9148921
  #> 4 184 19.5 0.8457030 Sat 1.666667 2010-07-03 19:00:00 0.8982188
  #> 5 184 20.0 0.8464793 Sat 1.943989 2010-07-03 19:30:00 0.8942396
  #> 6 184 20.5 0.8471040 Sat 2.679803 2010-07-03 20:00:00 0.8818705
$\end{example}
and the definitions of the variables are given as follows
\begin{itemize}
    \item \code{doy}: the day of the year, from 1 to 365;
    \item \code{tod}: the time of day, ranging from 18 to 22, where 18 indicates the period from 17:00 
    to 17:30, 18.5 the period from 17:30 to 18:00 and so on;
    \item \code{dem}: the demand (in kWh) during a 30min period, averaged over the 247 households;
    \item \code{dow}: factor variable indicating the day of the week;
    \item \code{temp}: the external temperature at Sydney airport, in degrees Celsius;
    \item \code{date}: local date and time;
    \item \code{dem48}: the lagged mean demand, that is the average demand (\code{dem}) during the same
    30min period of the previous day.
\end{itemize}
In particular, we are interested in analyzing the effect of temperature and time on the average demand distribution and its different quantiles. Hence the response variable is chosen to be \code{dem}, and the set of covariates contains \code{doy}, \code{tod}, \code{temp} and dummy variables representing \code{dow}. 
\begin{example}
  library(tidyr)
  ## Encode dow using dummy variables
  meanDem <- meanDem %>% mutate(value=1) %>% spread(dow, value, fill=0)

  Y <- meanDem$dem
  X <- as.matrix(meanDem %>% select(-c(date,dem,dem48)))
$\end{example}
Given that the sample size is fairly large, we consider fitting SPQR using the MLE method. The modeling parameters are selected using a grid search. Specifically, we select the number of basis functions (\code{n.knots}) from $\{10,15,20\}$ and the number of hidden neurons (\code{n.hidden}) from $\{10,15,20\}$. We focus on 2-hidden-layer neural networks so \code{n.hidden} represents the number of neurons in each hidden layer. For the control hyperparameters, we also select learning rate (\code{lr}) from $\{e^{-3},e^{-4},e^{-5}\}$; we set the batch size (\code{batchsize}) to be 128, maximum number of epochs (\code{epochs}) to be 800, and early stopping criterion (\code{early.stopping.epochs}) to be 50. The best model configuration is selected based on 10-fold CV error. 

\begin{figure}[htbp]
  \centering
  \includegraphics[width=\linewidth]{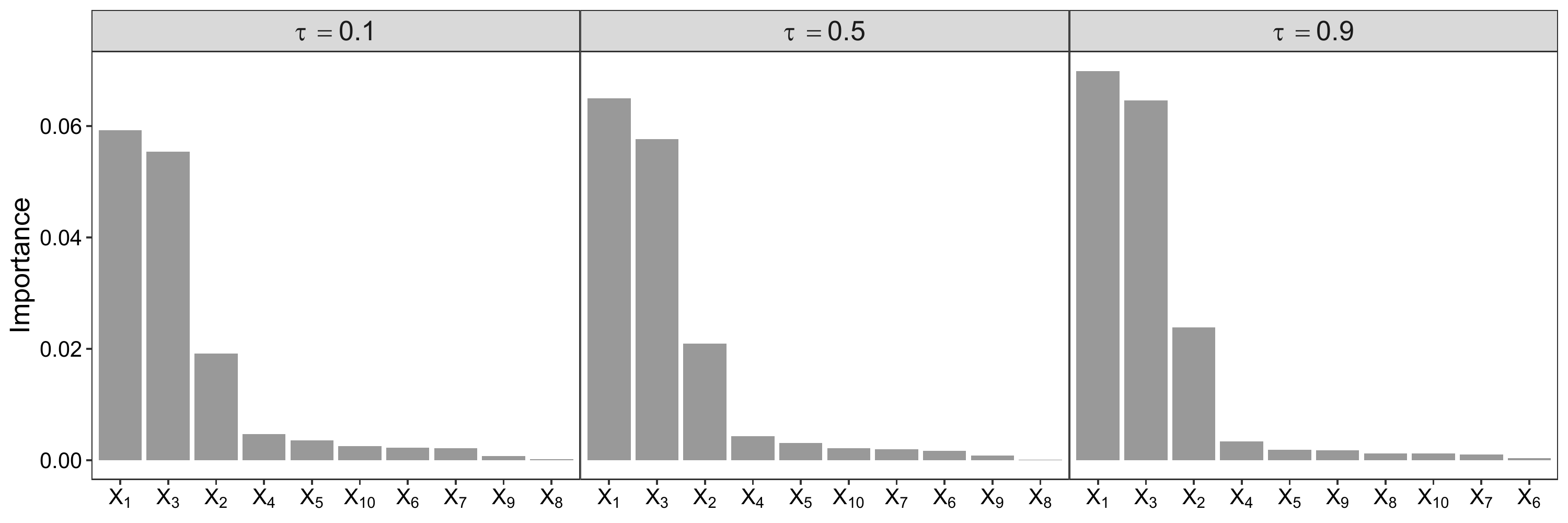}
  \caption{Variable importance (VI) of Australia electricity demand data. Estimated quantile VI of all covariates at $\tau\in\{0.1,0.5,0.9\}$.}
  \label{f:appvi}
\end{figure}

To accelerate the grid search routine, the function \code{foreach()} from the \CRANpkg{foreach} package \citep{foreach} can be used to run jobs in parallel. However, since tensor computation in the \pkg{torch} package is very memory-consuming, using many cores can easily exceeds the memory limit. In general, we recommend the user to start with 2 cores and then adjust accordingly to the available memory. The code snippet below gives an example of running the grid search routine over 3 cores.
\begin{example}
  library(foreach)
  library(doParallel)
  
  ## Fixed control hyperparams
  mle.control <- list(batch.size = 128, epochs = 800, 
                    use.GPU = TRUE, early.stopping.epochs = 50)
  tune.grid   <- expand.grid(lr = exp(-3:-5), n.knots = c(10,15,20), 
                             n.hidden = c(10,15,20))
  set.seed(919)
  ## Compute CV folds
  folds <- createFolds.SPQR(Y, nfold = 10)
  
  cl <- makeCluster(3)
  registerDoParallel(cl)
  ## 10-fold CV for model selection
  out <- foreach(i=1:nrow(tune.grid), .packages = c("SPQR","torch"), 
                 .combine = "rbind") %dopar% {
    lr <- tune.grid$lr[i]
    mle.control$lr <- lr
    n.knots  <- tune.grid$n.knots[i]
    n.hidden <- rep(tune.grid$n.hidden[i],2)
    mle.cv   <- cv.SPQR(X = X, Y = Y, folds = folds, n.knots = n.knots, 
                        n.hidden = n.hidden, method = "MLE", 
                        control = mle.control, normalize = TRUE,
                        verbose = FALSE, seed = 919)
    c(lr,n.knots,n.hidden,mle.cv$cve)
  }
  stopCluster(cl)
  
  ## Print the best model configuration
  cve <- out[,5]
  best_param <- out[which.min(cve),]
  names(best_param) <- c("lr","n.knots","n.hidden","n.hidden","cve")
  best_param
  
  #>           lr       n.knots      n.hidden      n.hidden           cve
  #> 6.737947e-03            20            15            15     -161.6610 
$\end{example}
We refit the model with the best configuration, using 10\% of data as validation set for early stopping criterion. 
\begin{example}
  mle.control$valid.pct <- 0.1
  mle.control$lr <- exp(-5)
  mle.fit <- SPQR(X = X, Y = Y, n.knots = 20, n.hidden = c(15,15),
                  method = "MLE", control = mle.control, normalize = TRUE,
                  verbose = TRUE, seed = 919)
  ## Print model summary 
  print(mle.fit, showModel=TRUE)

  #> SPQR fitted using MLE approach
  #> 
  #> Learning rate: 0.006737947
  #> Batch size: 128
  #> 
  #> Model specification:
  #>     Layers
  #>    Input Output Activation
  #>       10     15       tanh
  #>       15     15       tanh
  #>       15     20    softmax
  #> 
  #> Loss:
  #>   train = -219.1049,  validation = -169.3304
  #> 
  #> Elapsed time: 6.10 minutes
\end{example}
To determine which covariates are important in predicting quantiles of \code{dem}, we compare the quantile VI of all covariates at $\tau\in\{0.1,0.5,0.9\}$. The result is shown in Figure~\ref{f:appvi}.
\begin{example}
  plotQVI(mle.fit, var.names = colnames(X), tau=c(0.1,0.5,0.9))
\end{example}
Across all three quantiles, \code{doy}, \code{temp} and \code{tod} have the highest VI, suggesting strong seasonal and hourly effects. These effects are also more prominent on upper tail of the average demand distribution than on central region and lower tail. While one might expect that there is also a daily/weekend effect on electric consumption, the estimated effect of \code{dow} seems to be almost negligible, suggesting no particular difference in terms of electric consumption across the week. One reason could be that only data between 17:30 and 21:30 of the day are used, in which case the probability of people being home is similar between weekdays and weekends. 

\begin{figure}[htbp]
  \centering
  \includegraphics[width=\linewidth]{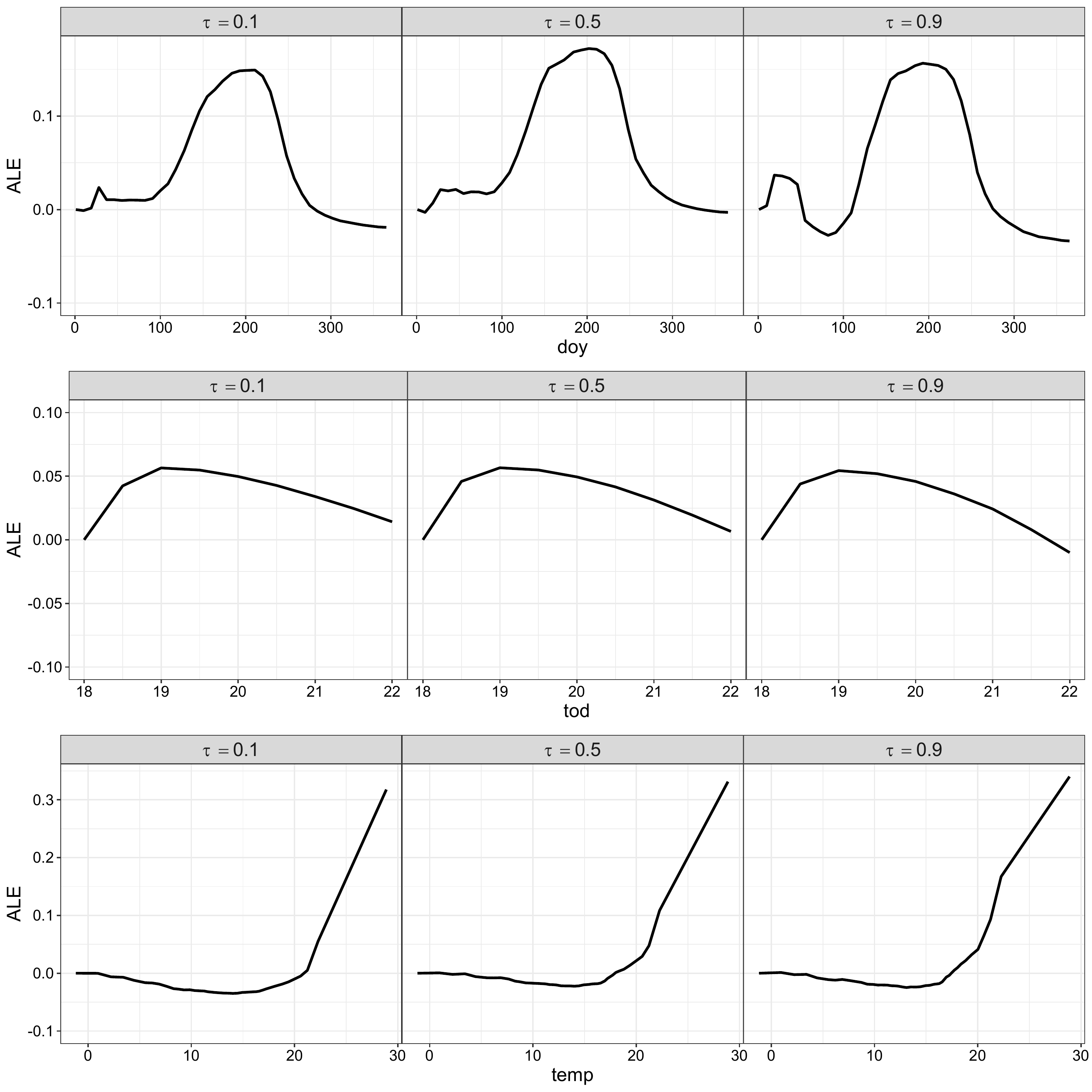}
  \caption{Estimated quantile ALE main effects of day of the year (\code{doy}), time of the day (\code{tod}) and temperature (\code{temp}) at $\tau\in\{0.1,0.5,0.9\}$.}
  \label{f:appale}
\end{figure}

To further investigate the effects of \code{doy}, \code{temp} and \code{tod} on the average demand distribution, we compute their quantile ALE main effects at $\tau\in\{0.1,0.5,0.9\}$. The results are plotted in Figure~\ref{f:appale}. 
\begin{example}
  for (j in 1:3) plotQALE(mle.fit, var.index=j, tau=c(0.1,0.5,0.9))
\end{example}
The estimated effect of \code{doy} has a unimodal shape at $\tau=0.1$ and $\tau=0.5$ and peaks at austral winter. Suggesting a high electric consumption when the weather is at its coldest. For $\tau=0.9$, the effect also has a minor mode at austral summer, suggesting that the average demand distribution is highly right-skewed during austral summer. The estimated effect of \code{tod} displays a quadratic upwarding trend before 20:00 and a linear downwarding trend afterwards. This suggests that use of electricity becomes more active as people start to get home and becomes less active as the night closes in. The downwarding trend is also more prominent at upper tail of the average demand distribution than at central region or lower tail. This suggests that as the day ends and people start to go to bed, the probability of very high electric consumption significantly decreases. The estimated effect of \code{temp} seems to complement that of \code{doy}. It has a "check" shape and shows a significant upwarding trend when temperature is above 20 degree. This suggests a significantly higher electric consumption when cooling becomes necessary during summer. We also notice that the turning point for \code{temp} effect at upper quantiles is closer to 15 degree than at lower quantiles. This may suggest that people who contribute to very high electric consumption tend to turn on cooling at a lower temperature than those who contribute to very low electric consumption. Similar characteristics of these effects were also observed in \citet{qgam}. 

\section{Summary}
In this article, we present the R package \pkg{SPQR} for fitting the semi-parametric conditional density and quantile regression models as proposed in \citet{xu2021bayesian}. The main advantage of these models is their capability of modeling complex covariate effects on the response distribution, which is absent in existing parametric quantile regression models. Furthermore, the estimated distribution and quantile functions are always valid, ensuring sensible inference even when the sample size is small. The package also provides a framework for fully Bayesian inference of the fitted models to allow uncertainty quantification, as well as model agnostic tools to understand the effects of different covariates on different parts of the distribution when model transparency is needed. We hope that this package will be a valuable addition to the existing family of quantile regression tools and be especially useful 
for analyzing complex and possibly high-dimensional heteroscedastic data where flexible density and quantile regression models are suitable. Future work will aim at improving the scalability of the Bayesian SPQR fitting framework by enabling sparsity-induced priors and implementing variational Bayes estimator. We also plan to include functions for more specialized use case, such as estimation of counterfactual distribution and quantile treatment effects in a causal inference context.

\bibliography{SPQR}

\section*{Acknowledgements}

This work was supported by grants from the National Science Foundation (DMS2152887, CBET2151651), the National Institutes of Health (R01ES031651-01), and the Southeast National Synthesis Wildfire and the United States Geological Survey's National Climate Adaptation Science Center (G21AC10045).

\address{Steven G. Xu\\
  North Carolina State University\\
  Raleigh, NC\\
  USA\\
  \email{sgxu@ncsu.edu}}

\address{Reetam Majumder\\
  North Carolina State University\\
  Raleigh, NC\\
  USA\\
  \email{rmajumd3@ncsu.edu}}

\address{Brian J. Reich\\
  North Carolina State University\\
  Raleigh, NC\\
  USA\\
  \email{bjreich@ncsu.edu}}

\end{article}

\end{document}